\documentclass[a4paper]{article}
\usepackage{amsmath,amssymb,graphicx,multirow}
\pdfoutput=1 

\usepackage{jheppub} 

\usepackage[T1]{fontenc} 

\makeatletter
\newsavebox{\@brx}
\newcommand{\llangle}[1][]{\savebox{\@brx}{\(\m@th{#1\langle}\)}%
  \mathopen{\copy\@brx\kern-0.5\wd\@brx\usebox{\@brx}}}
\newcommand{\rrangle}[1][]{\savebox{\@brx}{\(\m@th{#1\rangle}\)}%
  \mathclose{\copy\@brx\kern-0.5\wd\@brx\usebox{\@brx}}}
\makeatother

\newcommand{\mS}{\mu_\ell}
\newcommand{\tr}{\text{tr}}

\newcommand{\mT}{\left(\frac{\mS}{T}\right)}
\newcommand{\Ns}{N_\sigma}
\newcommand{\Nt}{N_\tau}

\newcommand{\M}{\mathcal{M}}

\renewcommand{\O}{\mathcal{O}}

\newcommand{\Z}{\mathcal{Z}}

\newcommand{\x}{\boldsymbol{x}}
\newcommand{\y}{\boldsymbol{y}}

\newcommand{\Tr}{\mathrm{Tr}}
\newcommand{\G}{\mathcal{G}}

\renewcommand{\Re}{\mbox{\textrm{Re}}}
\renewcommand{\Im}{\mbox{\textrm{Im}}}
\newcommand{\Tpc}{T_{pc}}

\newcommand{\hm}{\hat{\mu}_\ell}
\newcommand{\z}{\hat{z}}

\title{Pion screening mass at finite chemical potential}
\date{\today}

\author{Rishabh Thakkar}
\author{and Prasad Hegde}
\affiliation{Centre for High Energy Physics,\\Indian Institute of Science, Bangalore 560012, India.}

\emailAdd{rishabht@iisc.ac.in}
\emailAdd{prasadhegde@iisc.ac.in}

\abstract{
We present a method to compute the responses of meson screening masses to the chemical potential by Taylor expanding the correlator using lattice QCD simulation.  We start by comparing the free theory lattice results with the analytical expression. Then, using symmetry arguments, we obtain an expression for the correlator in a series of the chemical potential at finite temperature. Using this, we obtain the lowest order correction to the screening mass at a finite chemical potential for temperatures around 2.5 GeV. Our lattice analysis is limited to isoscalar chemical potential for the pseudoscalar channel. The calculations were performed using (2+1)-flavors of the Highly Improved Staggered Quark (HISQ/tree) action, with the ratio of the strange quark mass to the light quark mass $m_s/m_\ell=20$ corresponding to pion masses of 160 MeV.}

\begin{document}
\maketitle
\flushbottom

\section{Introduction}
\label{sec:introduction}
It is well-known that strongly-interacting nuclear matter undergoes a phase transition at high temperatures to a new state of matter called the quark-gluon plasma in which quarks and gluons are not confined within hadrons but are free to move throughout the volume of the system. This deconfinement is accompanied by the restoration of the chiral symmetry that is spontaneously broken at zero temperature. The nature of the phase transition depends upon the number of light quarks and their masses. For 2+1-flavor QCD with physical quark masses, the transition is known to be a crossover~\cite{Aoki:2006we} with a pseudocritical temperature  $\Tpc=156.5\pm1.5$~MeV~\cite{bazavov2019chiral}.

The properties of the quark-gluon plasma (QGP) have been studied extensively using a variety of approaches. Besides being of theoretical interest, an additional impetus for its study is provided by the various experiments in which the QGP is created in collisions of heavy nuclei at ultra-relativistic energies. The experimental results indicate that the QGP created in these experiments is strongly coupled~\cite{PHENIX:2004vcz,ALICE:2010suc,CMS:2012gvv}, which makes a theoretical description of the system challenging. Furthermore, the usual approach to calculating observables in field theories, namely perturbation theory, breaks down for Yang-Mills theories at finite temperatures beyond $\O(g^6)$, where $g$ is the Yang-Mills coupling constant, due to the severity of the infrared divergences~\cite{Linde:1980ts,Gross:1980br}. Even at lower orders, the series is slow to converge except at very high temperatures and successive corrections can even differ in sign. This latter issue however can be addressed through the resummation of the QCD perturbation series. Two widely used resummation schemes are Hard Thermal Loop (HTL) QCD~\cite{Pisarski:1988vd, Braaten:1989mz, Braaten:1989kk, Braaten:1990az}, and dimensionally reduced QCD or EQCD~\cite{Appelquist:1981vg, Nadkarni:1982kb, Nadkarni:1988fh, Kajantie:1995dw, Braaten:1995jr, Kajantie:2000iz}. These approaches have resulted in the determination of the QCD Equation of State (QEOS) to $\O(g^6\ln g)$~\cite{Kajantie:2002wa}. Alternatively, one can calculate these observables directly from the underlying theory of QCD using first-principles numerical simulations. This approach, known as lattice QCD, is a non-perturbative approach as it does not require the QCD coupling to be small. It has yielded precise estimates of many properties of the QGP~\cite{Bazavov:2014pvz, Bazavov:2017dus, HotQCD:2018pds, Bazavov:2019www, Bazavov:2020bjn, Bollweg:2021vqf, Bollweg:2022fqq, Bollweg:2022rps}.

Apart from bulk observables such as the pressure or the energy density, which are defined via the QCD partition function and its derivatives, there are also the spectral properties of the QGP defined in terms of various real or imaginary time thermal correlation functions. The most familiar of these observables are the various hadron correlators, which are the imaginary time two-point functions of the familiar hadron creation/annihilation operator $J_H$. By projecting these functions to zero transverse momentum ($p_x=p_y=0$) and zero frequency ($\omega=0$) in Fourier space by integrating over $x$, $y$ and the imaginary time $\tau$, one obtains the well-known screening correlators $C_H(z,T)$ of the hadron $H$ at temperature $T$, defined as 
\begin{equation}
    C_H(z,T) = \int_0^{1/T} d\tau \int dx dy \,\big\langle J_H^\dagger(x,y,z,\tau) J^{\phantom{\dagger}}_H(0,0,0,0) \big\rangle,
\end{equation}
where $J_H(x,y,z,\tau)$ is the hadron operator and the angular brackets represent the thermal average. As the separation $z\to\infty$, $C_H(z)\to e^{-zM_H(T)}$, where $M_H(T)$ is the screening mass at temperature $T$. As $T\to0$, $M_H(T)$ approaches the mass of the corresponding hadron. However, $M_H(T)$ is non-zero even in the QGP phase, that is, even when the quarks and gluons are deconfined. The screening mass thus provides information about the degrees of freedom present in the QGP at high temperatures. Additionally, since the hadron operators form multiplets according to the symmetries of the QCD Lagrangian, the corresponding correlators also become degenerate when the corresponding symmetry is restored, e.g., chiral symmetry restoration at $T\sim \Tpc$ or effective $U(1)_A$ restoration~\cite{Bazavov:2019www}, or also in the case of the appearance of possible new emergent symmetries at high temperatures~\cite{Rohrhofer:2019qal, Rohrhofer:2019qwq, Glozman:2022lda}. 

Among the various hadrons, the screening masses corresponding to the flavor non-singlet mesons have been the most studied since their calculation does not require the evaluation of the computationally expensive disconnected diagrams. Continuum-extrapolated results for the masses of all the flavor-singlet spin-0 and spin-1 mesons formed out of the light and strange quarks, over a temperature range 130 MeV $\lesssim T \lesssim$ 1000 MeV have been recently published~\cite{Bazavov:2019www}. Similar results are also available for the charm quark mesons, although these results have not yet been continuum-extrapolated~\cite{Bazavov:2014cta}.

The above discussion assumed that the QGP is at zero quark chemical potential, $\mu_u=\mu_d=\mu_s=0$. Collisions at lower beam energies produce a QGP that is at non-zero baryochemical potential $\mu_B$ at freeze-out~\cite{Andronic:2005yp,Andronic:2017pug}. This makes it possible to also study the properties of the QGP in the $T$-$\mu_B$ plane. The phase diagram of QCD in the $T$-$\mu_B$ plane is a topic of great interest and various phases of nuclear matter have been conjectured~\cite{Halasz:1998qr, Rajagopal:1999cp, Rajagopal:2000wf, Stephanov:1998dy}. One such prediction is that the QCD chiral crossover transition turns into a first-order transition line at a second-order $Z(2)$ critical point. This is the famous conjectured QCD critical point. The change from a crossover to a genuine phase transition should have consequences for physical observables including screening masses and hence a knowledge of the screening masses at finite chemical potential should be able to provide some information regarding the existence and location of the QCD critical point. Unfortunately, lattice QCD calculations are not possible at finite chemical potential due to the infamous sign problem of lattice QCD. Although a complete solution to the sign problem is not known, several partial approaches have been proposed among which the method of Taylor expansions~\cite{Gavai:2003nn, Allton:2003vx} has also been applied to calculate second-order corrections to screening masses~\cite{Pushkina:2004wa,QCD-TARO:2001lhr} as well as temporal correlators~\cite{Nikolaev:2020vll}.

In this paper, we will present a new way of calculating the second Taylor coefficient $M''(0)/2$ of the screening mass with respect to the isoscalar chemical potential $\mS$, defined as $\mu_u=\mu_d=\mS$, $\mu_s=0$. Our approach derives from the exact result for the free theory screening correlator at finite $\mS$ presented in Ref.~\cite{vepsalainen2007mesonic}. We thus expect our approach to be reliable at high temperatures. We first calculate the free theory isoscalar screening correlator to $\O(\hm^4)$ using the Highly Improved Staggered Quark (HISQ) formulation on an $80^3\times8$ lattice and compare our results with the exact expressions. While we obtain good agreement with the theoretical expressions, we also find that we need to go to large $zT$ in order to achieve this agreement. Next, we repeat the calculation at finite temperature using $64^3\times8$ lattices at two temperatures viz. $T=2.24$ and 2.90 GeV. We find $M''(0)$ to be small but non-zero within error at the smaller temperature of $T=2.24$~GeV, while it is consistent with zero within error at the higher temperature of $T=2.90$~GeV. We expect these results to improve as the fit window is moved towards larger $zT$. It should therefore be possible to improve upon these estimates in the future by working with lattices having a larger aspect ratio.

Our paper is organized as follows: In \autoref{sec:scrcorr_finite_density}, we outline the calculation of the pseudoscalar screening correlator and its Taylor coefficients in lattice QCD starting from the QCD partition function at non-zero $T$ and $\mS$. The exact form of the correlator is known in the continuum for the free theory with massless quarks. Using our formalism, we calculate the free theory screening correlator and its Taylor coefficients up to the fourth order on the lattice and compare our results with the corresponding continuum expressions in \autoref{sec:free_theory}. We repeat the same calculation in \autoref{sec:finite_T}, but this time for two finite temperatures in the range $T \sim 2$ - 3 GeV. We present an \emph{ansatz} motivated by the free theory expression and compare it with the obtained results. We also describe a procedure for extracting the $\O(\mS^2)$ correction to the $\mS=0$ pseudoscalar screening mass using that ansatz above and present our results for the two above temperatures. We state our conclusions in \autoref{sec:conclusions}. We present the formulas for the screening correlator and its first four Taylor coefficients in terms of the derivatives of the fermion propagator and fermion determinant in \autoref{app:corr_derivatives}. In \autoref{app:isoscalar_chempot}, we present the different operators that are required for calculating the derivatives of the screening correlator.

\section{Screening Correlators at Finite Density}
\label{sec:scrcorr_finite_density}
We consider lattice QCD with 2+1 flavors of staggered quarks on an $\Ns^3\times\Nt$ lattice in Euclidean spacetime. The partition function at finite temperature $T$ and isoscalar chemical potential $\mS$ is given by
\begin{equation}
    \Z(T,\mS) = \int \mathcal{D}U\,\Delta(T,\mS)\,e^{-S_G(T)},
\label{eq:partition_function}
\end{equation}
where the integral is over all gauge links $U$, $S_G$ is the gauge action, and $\Delta(T,\mu_\ell)$ is the fermion determinant given by
\begin{equation}
\Delta(T,\mS) = \prod_{f=u,d,s} \big[\det M_f(m_f,T,\mu_f)\big]^{1/4},
\label{eq:determinant}
\end{equation}
where $M_f(m_f,T,\mu_f)$ is the staggered fermion matrix for flavor $f$. In the present case, we have considered $m_u=m_d=m_\ell=m_s/20$, and $\mu_u=\mu_d=\mS$, $\mu_s=0$.

A staggered meson operator is given by $\M(\x) \equiv \sum_{\x'}\bar{\chi}_i(\x)\,\phi(\x,\x')\,\chi_j(\x')$, where $\x$ and $\x'$ are sites belonging to the same unit hypercube, $\bar{\chi}_i$ and $\chi_j$ are staggered quark fields with flavor indices $i$ and $j$ respectively, and $\phi(\x,\x')$ is a phase factor that depends upon the spin and taste quantum numbers of the meson~\cite{Altmeyer:1992dd}. For a \emph{local} meson operator, the phase factor is given by $\phi(\x,\x')=\phi(\x)\delta_{\x,\x'}$ and the meson operator then simply becomes $\M(\x) = \phi(\x)\bar{\chi}_i(\x)\chi_j(\x)$.

The finite-$\mS$ meson correlator $\G(\x,T,\mS)$ is the two-point function of the corresponding meson operator: $\G(\x,T,\mS) \equiv \llangle \M(\x) \overline{\M}(0) \rrangle$, where $\x = (x,y,z,\tau)$ and the double angular brackets $\llangle \cdot \rrangle$ denote a thermal expectation value at $\mS\ne0$ viz.
\begin{equation}
    \llangle[\big] \O(\mS) \rrangle[\big] = \frac{1}{\Z(T,\mS)}\int \mathcal{D} U \, e^{-S_G(T)} \, \O(\mS)\,\Delta(T,\mS).
    \label{eq:angular_brackets-i}
\end{equation}

For the rest of this paper, we shall only consider the two-point function of the staggered light pseudoscalar meson, for which $(i,j)=(u,d)$ and $\phi(\x)=1$ for all $\x$. The expectation value $\llangle \M(\x) \overline{\M}(0) \rrangle$ can then be shown to be~\cite{cheng2011meson}
\begin{equation}
    \llangle[\big] \M(\x) \overline{\M}(0) \rrangle[\big] =
    \llangle[\big] \tr\big[P_u(\x,0,\mu_u) P_d^\dagger(\x,0,-\mu_d)\big] \rrangle[\big],
\end{equation}
where the trace is over the color indices and $P_k(\x,\y,\mu_k)$ is the staggered quark propagator for the $k$th flavor. We can drop the flavor indices $u$ and $d$ since the up and down quarks are identical in the 2+1 flavor case. Setting $\mu_u=\mu_d\equiv\mS$ in the above equation, and denoting 
\begin{equation}
\tr \big[P(\x,0,\mS) P^\dagger(\x,0,-\mS)\big] \equiv G(\x,\mS),
\label{eq:meson_operator}
\end{equation}
we can write
\begin{equation}
    \G(\x,T,\mS) = \llangle[\big] G(\x,\mS) \rrangle[\big] =
    \frac{\int \mathcal{D} U \, e^{-S_G(T)} \, G(\x,\mS)\,\Delta(T,\mS)}{\int \mathcal{D} U \, e^{-S_G(T)} \, \Delta(T,\mS)}.
\label{eq:pion_corr}
\end{equation}
Owing to the sign problem of lattice QCD, it is not possible to calculate $\G(\x,T,\mS)$ directly. Instead, we expand $\G(\x,T,\mS)$ in a Taylor series in $\mS/T$:
\begin{equation}
    \G(\x,T,\mS) = \sum_{k=0}^\infty \frac{\G^{(k)}(\x,T)}{k!}\mT^{k},
\label{eq:taylor_expansion}
\end{equation}
where the Taylor coefficients $\G^{(k)}(\x,T)$ are evaluated at $\mS=0$. By differentiating \autoref{eq:pion_corr} w.r.t. $\mS/T$, we find that the first three Taylor coefficients are given by
\begin{align}
    \G^{(0)}(\x,T) &= \langle G \rangle, \notag \\ \G^{(1)}(\x,T) &= \langle G' \rangle,\;\text{and} \notag \\
    \G^{(2)}(\x,T) &= \left\langle  {G}'' + 2{G}'\, \frac{ {\Delta}' }{ \Delta }
+G \,\frac{{\Delta}'' }{ \Delta } \right\rangle
-  \left\langle G \right\rangle  \left\langle \frac{ {\Delta}'' }{ \Delta } \right\rangle,
\label{eq:Ck_zero_to_two}
\end{align}
where the primes denote differentiation w.r.t. $\hm\equiv\mS/T$ and the single angular brackets $\langle\cdot\rangle$ denote $\mS=0$ thermal expectation values, {\it i.e.},
\begin{equation}
    \big\langle \O(0) \big\rangle = \frac{1}{\Z(T,0)}\int \mathcal{D} U\,e^{-S_G(T)}\,\O(0)\,\Delta(T,0).
    \label{eq:angular_brackets-ii}
\end{equation}
We have dropped terms containing the expectation value of odd derivatives of the determinant $\Delta$ in \autoref{eq:Ck_zero_to_two} since it can be shown that they vanish at $\mS=0$~\cite{gottlieb1988fermion}. For the present work, we also require the Taylor coefficients for third and fourth orders, and hence we give the corresponding expressions in \autoref{app:corr_derivatives}. In \autoref{app:isoscalar_chempot}, we also give the various operator equations required for the calculation of the terms in \autoref{eq:Ck_zero_to_two} and \autoref{app:corr_derivatives}.

The screening correlator $C(z,T,\mS)$ is obtained from $\G(\x,T,\mS)$ by summing over $x$, $y$ and $\tau$ i.e.
\begin{equation}
    C(z,T,\mS) = \frac{1}{\Nt\Ns^2}\,\sum_{x,y,\tau} \G(\x,T,\mS).
\label{eq:scr_corr}
\end{equation}
Its Taylor expansion follows straightforwardly from \autoref{eq:taylor_expansion}, namely
\begin{align}
    C(z,T,\mS) &= \sum_{k=0}^\infty \frac{C^{(k)}(z,T)}{k!}\mT^{k}, \notag \\
    C^{(k)}(z,T) &= \frac{1}{\Nt\Ns^2} \sum_{x,y,\tau} \G^{(k)}(\x,T).
    \label{eq:sum_corr}
\end{align}

\section{Free  Theory Screening Correlator at $\boldmath{\mS\ne0}$}
\label{sec:free_theory}
\autoref{eq:scr_corr} can be calculated exactly for free massless quarks in continuum QCD. For $zT \gg 1$, the screening correlator is given by~\cite{vepsalainen2007mesonic}
\begin{equation}
    \frac{C_\text{free}(z,T,\mS)}{T^3} = \frac{3}{2}\frac{e^{-2\pi zT}}{zT}
    \left[\left(1+\frac{1}{2\pi zT}\right)\cos(2z\mS) + \frac{\mS}{\pi T}\sin(2z\mS)\right] + \O\left(e^{-4\pi zT}\right).
\label{eq:scr_corr_free}
\end{equation}
By differentiating w.r.t. $\hm$, we obtain the first few Taylor coefficients as (with $\z\equiv zT$)
\begin{align}
&\frac{C^{(0)}_\text{free}(z,T)}{T^3} = \frac{3e^{-2\pi \z}}{2\z}\left(1+\frac{1}{2\pi \z}\right),& &&
\frac{C^{(2)}_\text{free}(z,T)}{T^3} &= 6\z e^{-2\pi \z}\left(\frac{1}{2\pi\z}-1\right), && \notag \\
&\frac{C^{(4)}_\text{free}(z,T)}{T^3} = 24\z^3e^{-2\pi \z}\left(1-\frac{3}{2\pi \z}\right),& && C^{(1)}_\text{free}(z,T) &= C^{(3)}_\text{free}(z,T)=0. &&
\label{eq:taycoeff_free}
\end{align}
The odd-numbered Taylor coefficients are identically zero while the even-numbered Taylor coefficients are non-zero and share the same exponential decay factor. Removing the exponential decay of the correlator, we define the amplitude
\begin{align}
    A_\text{free}\equiv \left(\frac{C_\text{free}}{T^3}\right)\z\,e^{2\pi \z}= \frac{3}{2}    \left[\left(1+\frac{1}{2\pi zT}\right)\cos(2z\mS) + \frac{\mS}{\pi T}\sin(2z\mS)\right]
    \label{eq:amp}
\end{align}
Similar to the correlator, we can expand the amplitude in a Taylor series
\begin{align}
    A(z,T,\mS) &= \sum_{k=0}^\infty \frac{A^{(k)}(z,T)}{k!}\mT^{k}, 
    \label{eq:sum_amp}
\end{align}
where $A^{(k)}$ are the Taylor coefficients for the amplitude obtained by taking the derivatives of \autoref{eq:amp}.
We also define the ratios
\begin{align}
&& && \Gamma(\z) \equiv \frac{C^{(2)}(z,T)}{C^{(0)}(z,T)} && \text{and} &&
      \Sigma(\z) &\equiv \frac{C^{(4)}(z,T)}{C^{(0)}(z,T)}\,, && &&
\end{align}
which gets rid of the exponential factor and in the large-$\z$ limit, we obtain:
\begin{align}
\Gamma_\text{free}(\z) &=-4\z^2\left(1-\frac{1}{2\pi\z}\right)\Big/\left(1+\frac{1}{2\pi\z}\right),& &&
\Sigma_\text{free}(\z) &= 16\z^4\left(1-\frac{3}{2\pi\z}\right)\Big/\left(1+\frac{1}{2\pi\z}\right),& \notag \\
&= - 4\z^2 + \frac{4\z}{\pi} -\frac{2}{\pi^2}+\O\left(\z^{-1}\right),&  &&
&= 16\z^4 - \frac{32\z^3}{\pi} + \frac{16\z^2}{\pi^2}+\O\left(\z\right),& \notag \\
&\equiv \alpha_2\z^2+\alpha_1\z+\alpha_0,& && &\equiv \beta_4\z^4+\beta_3\z^3+\beta_2\z^2.&
\label{eq:gamma_sigma_free}
\end{align}

\par The above equations provide the Taylor expansions for both $\Gamma_\text{free}(\z)$ and $\Sigma_\text{free}(\z)$ truncated at the fourth term. The Taylor expansion has coefficients with alternating signs. The truncated terms contribute less than 2\% and 4\% respectively for $\z > 1$ which become less significant with increasing $\z$. The expansion starts at $\O(\z^2)$ for $\Gamma_\text{free}(\z)$ and at $\O(\z^4)$ for $\Sigma_\text{free}(\z)$. In the large-$\z$ limit therefore, $\Gamma_\text{free}(\z)$ and $\Sigma_\text{free}(\z)$ are approximately given by quadratic and quartic polynomials respectively. We will see that this remains true when we generalize the free theory expressions to the finite temperature case in \autoref{sec:finite_T}.

To verify \autoref{eq:taycoeff_free} and \autoref{eq:gamma_sigma_free}, we calculated $C^{(0)}_\text{free}$, $C^{(2)}_\text{free}$ and $C^{(4)}_\text{free}$ using the HISQ/tree staggered quark action on an $80^3\times8$ lattice using a gauge configuration with gauge links equal to the unit $3\times3$ matrix. To ensure the convergence of the fermion matrix inverter, it was necessary to work with a non-zero light quark mass. However, we repeated our simulation with different quark masses $am_\ell=0.01$, 0.00014, and 0.00001, with the stopping residual always equal to $10^{-9}$, and found that the results we obtained were independent of the quark mass up to very small differences at large $\z$. We, therefore, felt confident that the results we had obtained were quite close to the results for free massless quarks.

\begin{figure}
    \begin{center}
    \includegraphics[width=.48\linewidth]{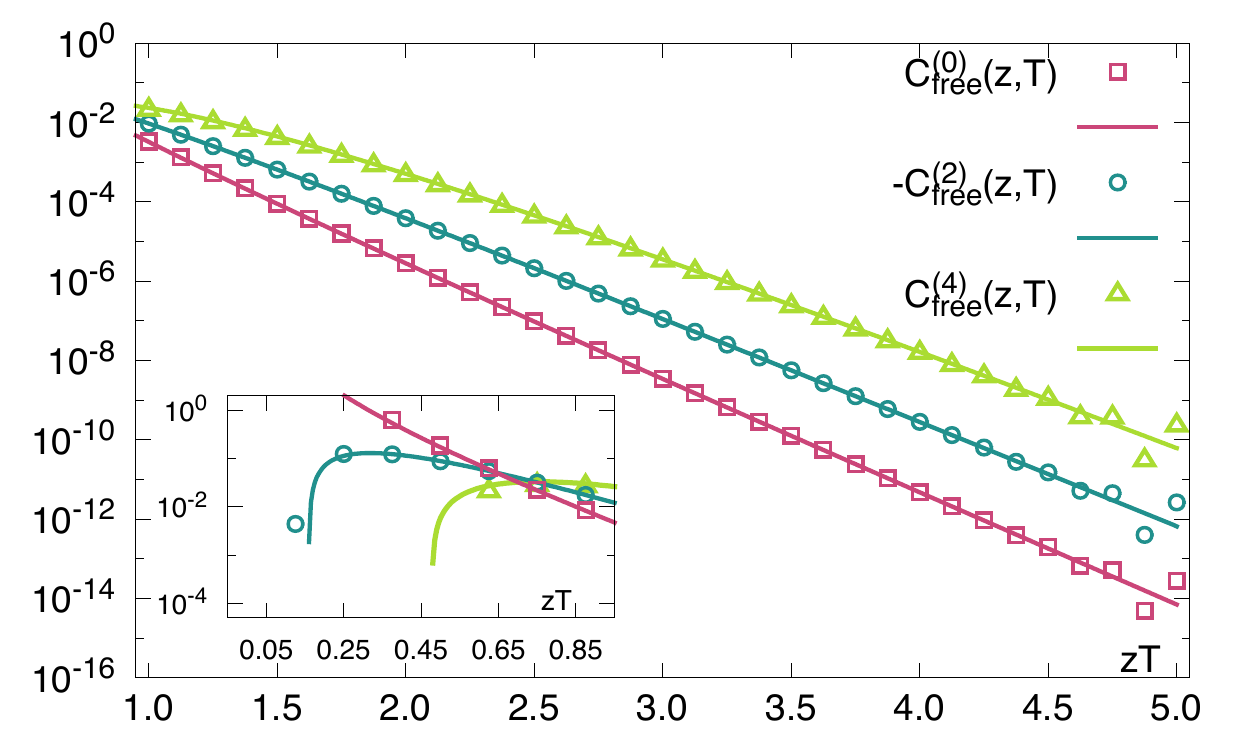}        
    \includegraphics[width=.48\linewidth]{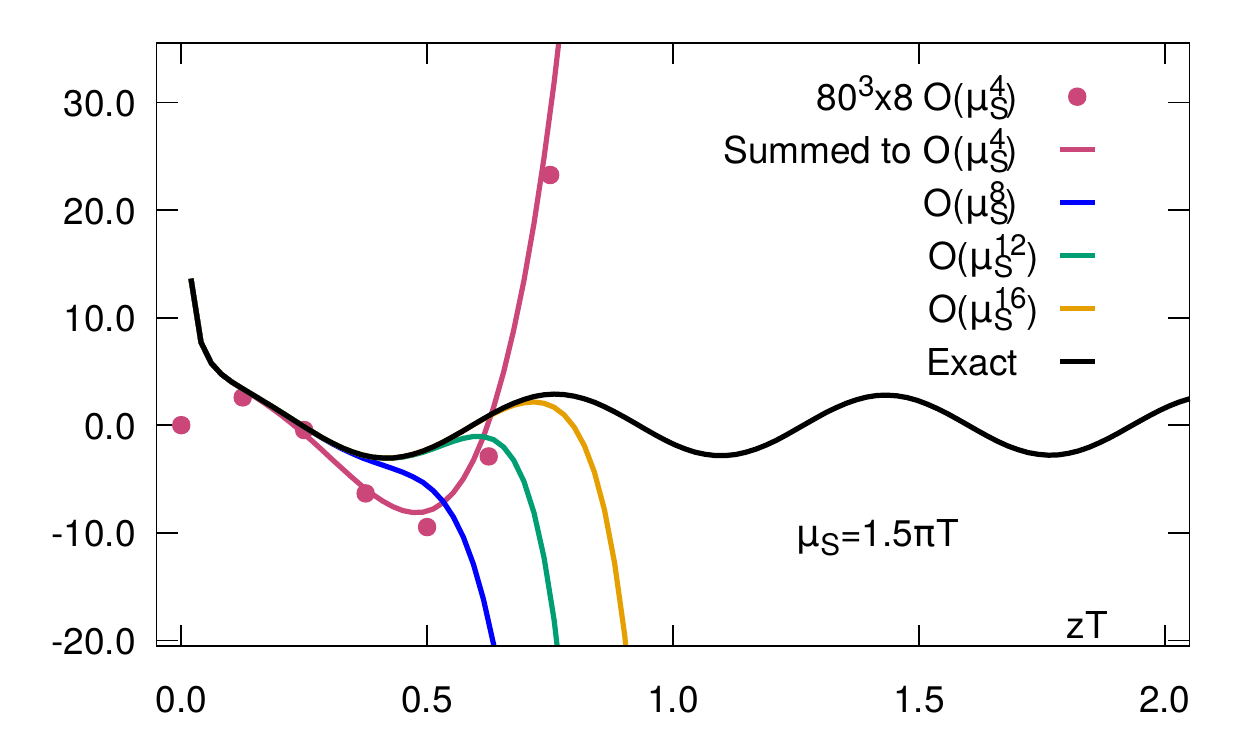}%
    \end{center}
\caption{\small (left) Lattice calculations of $C_\text{free}^{(0)}$, $C_\text{free}^{(2)}$ and $C_\text{free}^{(4)}$ compared to the corresponding free theory expressions obtained from \autoref{eq:taycoeff_free}. Points are lattice results while solid lines are the corresponding continuum expressions. The main plot shows the results for the range $1 \lesssim \z \leq 5$, while the results for $0 \leq \z \lesssim 1$ are plotted in the inset. (right) Summed amplitude for various orders of ($\hm)$ as well as the exact expression from \autoref{eq:amp} are plotted for $\hm=1.5\pi$. The lattice data summed up to $\mathcal{O}(\hm^4)$ is also plotted.}
\label{fig:free_CbyC_isoscalar}
\end{figure}

We plot our results for $C_\text{free}^{(0)}$, $C_\text{free}^{(2)}$ and $C_\text{free}^{(4)}$, as obtained for $am_\ell=0.00014$, in \autoref{fig:free_CbyC_isoscalar} (left). In the same figure, we also compare our results with the corresponding theoretical expressions as given in \autoref{eq:taycoeff_free}. We plot our results as functions of $zT \equiv \z$ i.e. as functions of the separation $z$ in units of the inverse temperature $T^{-1}$. Due to the use of periodic boundary conditions in the simulation, the largest $z$ value possible was $z_\text{max}= \Ns a/2 = 40a$, where $a$ was the lattice spacing and $\Ns$ was the number of sites in the $z$ direction. This maximum separation was equivalent to $\z_\text{max} = 5$ since $T=1/a\Nt$ and the number of sites $\Nt$ in the temporal direction was equal to 8.

Although \autoref{eq:taycoeff_free} are valid for $\z \gg 1$, we find that the theoretical curves agree well with the lattice data down to $\z \simeq 0.30$. Close to $\z=5$ on the other hand, we find that the lattice data oscillate about the corresponding theoretical curves. These oscillations, which vanish in the continuum limit, exist for all $\z$ and increase as $\z$ is increased. They are a well-known feature of both temporal as well as spatial correlators calculated for free staggered fermions and arise because the functional form of the staggered quark propagator is different for even and odd sites~\cite{Boyd:1992uk,Boyd:1994np}.

Using the obtained Taylor coefficients, the summed correlators were calculated using Eq. \eqref{eq:sum_corr}. To compare the summed correlator with the exact expression, the summed amplitude \autoref{eq:sum_amp} for various orders of $\hm$ are plotted in \autoref{fig:free_CbyC_isoscalar} (right) for $\hm=1.5\pi$. The summed amplitudes agree with the exact expression for a finite distance after which it diverges. This distance of agreement increases with increasing the number of summed terms. Like the exact expression, the summed amplitudes also display oscillatory behavior.
The summed amplitude using lattice data up to $\mathcal{O}(\hm^{16})$ is also plotted. The lattice data and analytic expression have a good agreement. Repeating the analysis with various values of $\hm$, similar behavior was observed. The summed amplitude with smaller values of $\hm$ has an agreement with the exact expression for larger values of $\z$ before it starts diverging.

\begin{figure}
    \includegraphics[width=.48\linewidth]{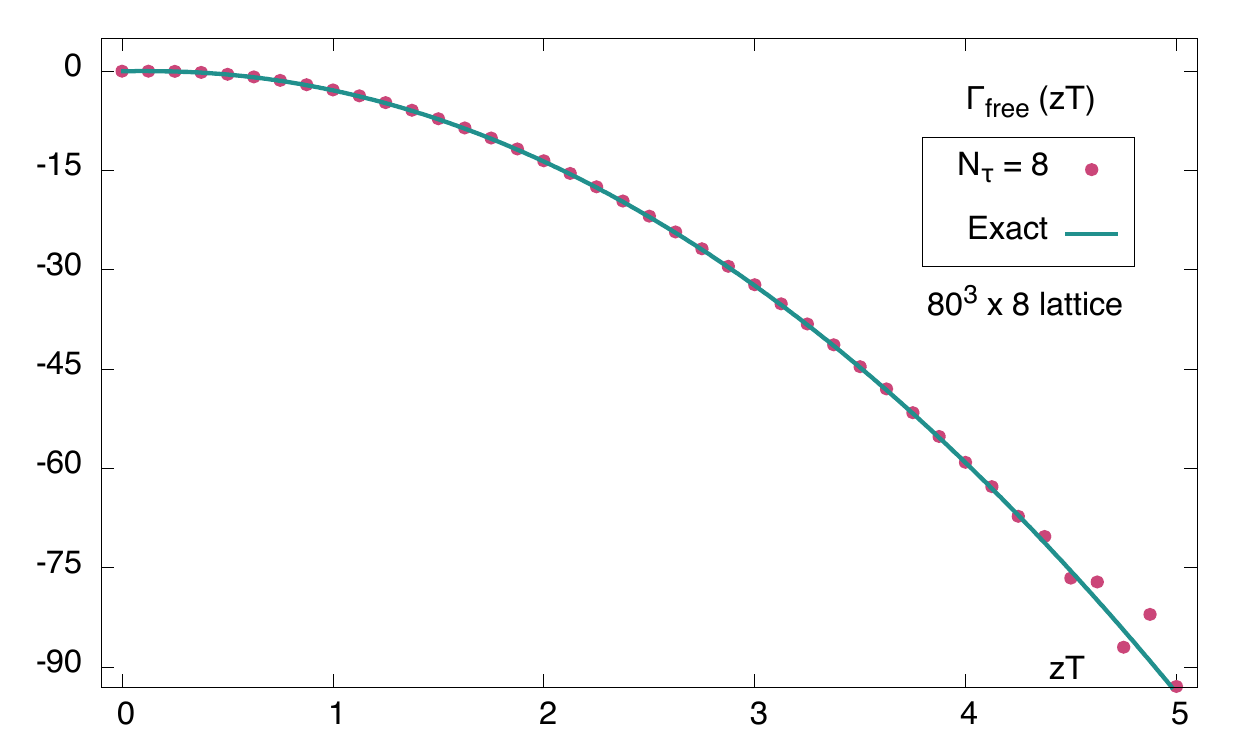}%
    \hspace{0.04\textwidth}%
    \includegraphics[width=.48\linewidth]{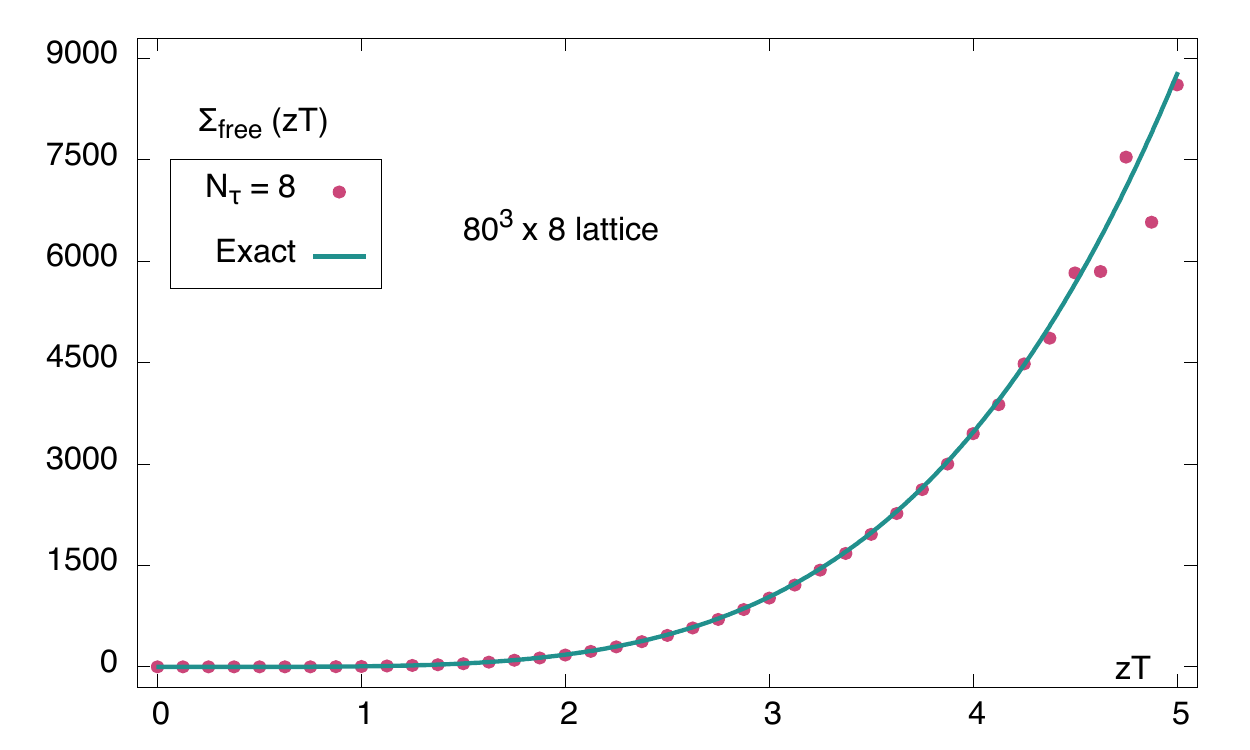}
\caption{\small Lattice calculations of $\Gamma_\text{free}$ (left) and $\Sigma_\text{free}$ (right) compared to the corresponding free theory expressions obtained from \autoref{eq:gamma_sigma_free}. Points are lattice results while solid lines are the continuum results.}
\label{fig:free_GS_isoscalar}
\end{figure}

In \autoref{fig:free_GS_isoscalar} (left) and \autoref{fig:free_GS_isoscalar} (right), we plot our results for $\Gamma_\text{free}(\z)$ and $\Sigma_\text{free}(\z)$ respectively and compare them with the corresponding theoretical expressions derived from~\autoref{eq:taycoeff_free}. Once again, we find very good agreement between our lattice data and the theoretical expressions, in fact seemingly right up to $\z=0$. The reason for this extended agreement is due to the fact that the ratios $\Gamma_\text{free}(\z)$ and $\Sigma_\text{free}(\z)$, unlike the Taylor coefficients themselves, remain finite in the $\z\to0$ limit.

Despite the good agreement between our results for $\Gamma_\text{free}(\z)$ and $\Sigma_\text{free}(\z)$ and the exact expressions, we also fit the data to polynomials $\alpha_2\z^2+\alpha_1\z+\alpha_0$ and $\beta_4\z^4+\beta_3\z^3+\beta_2\z^2$ and compared the fit results with the exact values given in~\autoref{eq:gamma_sigma_free}. In \autoref{sec:finite_T}, we will present a procedure in which the $\O(\hm^2)$ corrections to the $\mS=0$ screening mass and screening amplitude in the finite temperature case can be obtained from the coefficients of polynomial fits to the lattice data for $\Gamma(\z)$ and $\Sigma(\z)$. The polynomial \emph{ans\"atze} are valid only in the large-$\z$ limit. By fitting our free theory results to polynomial \emph{ans\"atze} for different fit ranges, we can obtain an idea for the minimum $\z$ range above which such fits may be carried out.

We present our results for the fits in \autoref{tab:fit_free}. The fits were carried out for two ranges 
$\left[\z_\text{min},\z_\text{max}\right] = [1.0,\,4.0]$ and $[2.0,\,4.0]$. The upper fit window of $\z=4$ was chosen to avoid the non-physical oscillation affecting the fit parameters. We note that the polynomials~\autoref{eq:gamma_sigma_free} are derived from \autoref{eq:taycoeff_free} as approximations that are valid in the large-$\z$ limit. We, therefore, expect the fit results to improve as $\z_\text{min}$ is increased. Conversely, by keeping more terms in \autoref{eq:gamma_sigma_free}, one should be able to fit the data over a wider $\left[\z_\text{min},\z_\text{max}\right]$ range.

\begin{table}[!b]
\hspace{-0.025\textwidth}%
\begin{tabular}{|c|c|c|c|c|c|c|} \hline
 Fit range & $-\alpha_2$ & $\alpha_1$ & $-\alpha_0$ & $\beta_4$ & $-\beta_3$ & $\beta_2$ \\ \hline
\multirow{2}{*}{$1.0\leq\z\leq4.0$} & 3.985(3)  & 1.20(1) &          & 15.97(5) & 10.21(19) & \\ \cline{2-7}
                                    & 4.018(6) & 1.37(3) & 0.20(4)   & 16.39(18) & 12.9(1.1) & 4.0(1.6) \\ \hline
\multirow{2}{*}{$2.0\leq\z\leq4.0$} & 3.995(4)  & 1.24(1)  & & 15.99(7) & 10.29(24) & \\ \cline{2-7}
                                    &  4.04(2)  &  1.53(11)  & 0.44(17) & 16.63(33) & 14.4(2.1) & 6.6(3.4)\\\hline
Exact & $4$ & $4/\pi\approx1.273$ & $2/\pi^2\approx0.203$ & 16 & $32/\pi\approx10.186$ & $16/\pi^2\approx1.621$\\ \hline 
\end{tabular}
\caption{Results of polynomial fits $\alpha_2\z^2+\alpha_1\z+\alpha_0$ and $\beta_4\z^4+\beta_3\z^3+\beta_2\z^2$ to $\Gamma_\text{free}(\z)$ and $\Sigma_\text{free}(\z)$ respectively, with and without the lowest order coefficients $\alpha_0$ and $\beta_2$, for the fit ranges $1.0\leq\z\leq3.0$ and $2.0\leq\z\leq4.0$.}
\label{tab:fit_free}
\end{table}

We present our fit results in \autoref{tab:fit_free}. The fits were carried out using~\autoref{eq:gamma_sigma_free}, both with and without the lowest order coefficients $\alpha_0$ and $\beta_2$. In each case, the fit was performed for two ranges in $\z$, namely $[1.0,\,4.0]$ and $[2.0,\,4.0]$. The obtained fit coefficients are fairly close to the expected free theory values. One expects the contribution of the lower order terms to become more important, and hence the coefficients $\alpha_0$ and $\beta_2$ to be better determined, as $\z_\text{min}$ is decreased. This is indeed what we observe from a comparison of the fit results in the two fit ranges. On the other hand, retaining these coefficients in the fit range $[2.0,\,4.0]$ yields poorer fits when compared with fits with these terms dropped.

We also note from \autoref{tab:fit_free} that the systematic error, namely the variation of the fit coefficients with the change in the fit range, can exceed the statistical error, which is the error on the fit coefficients themselves. For example, in the fit range $[1.0,\,4.0]$ and without the coefficient $\alpha_0$, one obtains $-\alpha_2=3.985(3)$ i.e. a result that is five standard deviations away from the true value of 4. When the fit range is changed to $[2.0,\,4.0]$, one obtains $-\alpha_2=3.995(4)$, which is in much better agreement with the true value. Thus, in order to determine the highest order coefficients, one must either retain sufficiently many lower order terms or go to large enough $\z_\text{min}$ that the contribution of the lower order terms can be neglected. We will keep this in mind when we fit the finite-temperature data in \autoref{sec:finite_T}.

\begin{figure}
    \includegraphics[width=.48\linewidth]{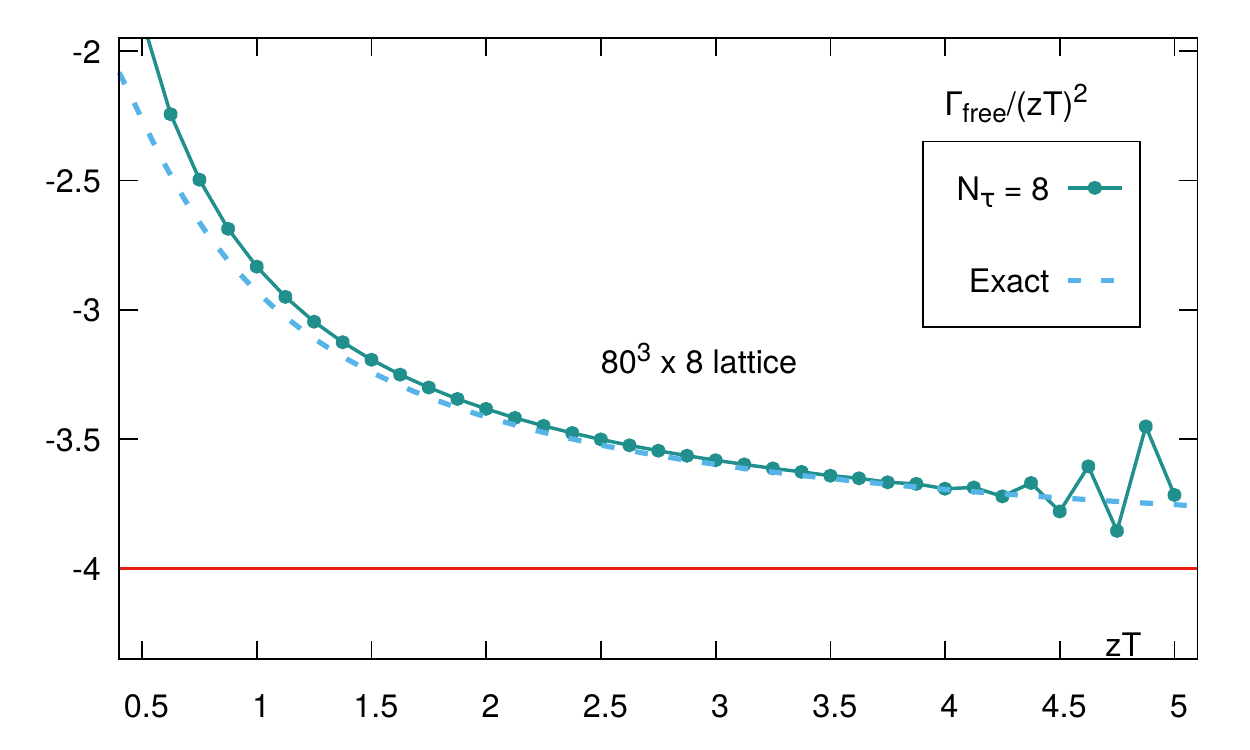}%
    \hspace{0.04\textwidth}%
    \includegraphics[width=.48\linewidth]{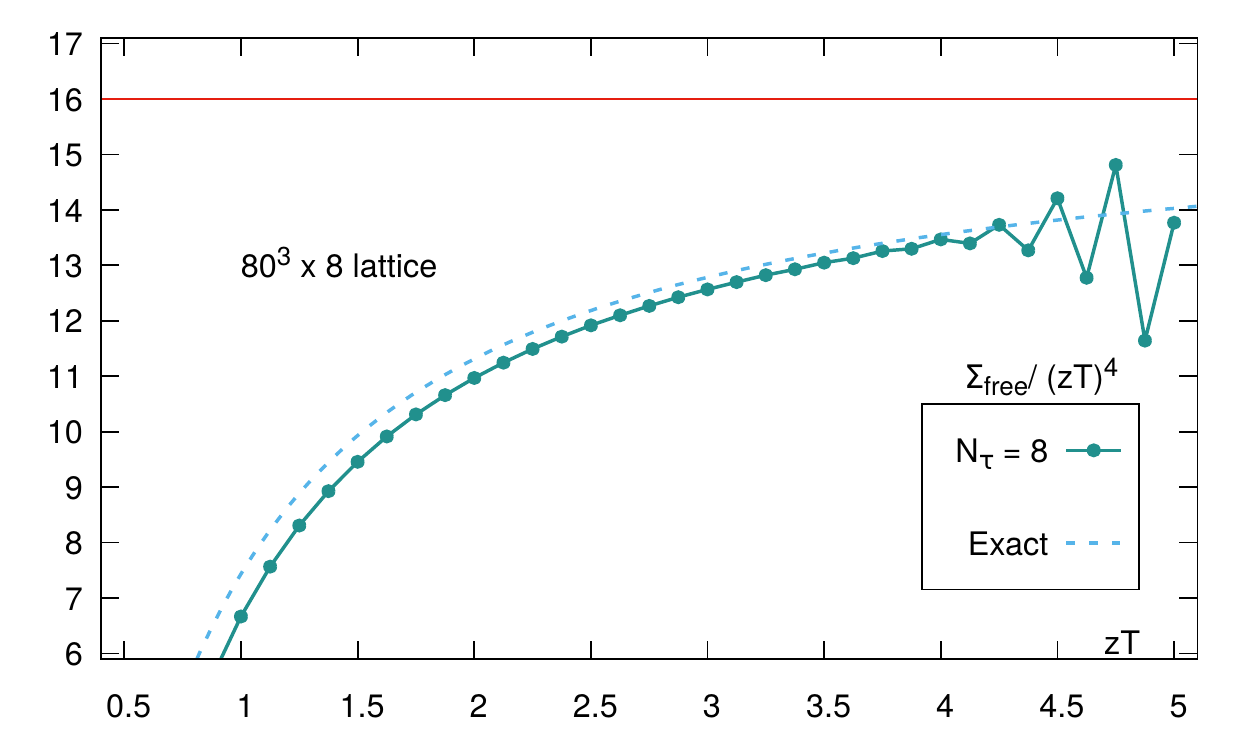}
\caption{\small Lattice calculations of $\Gamma_\text{free}/(zT)^2$ (left) and $\Sigma_\text{free}/(zT)^4$ (right), compared with the corresponding free theory expressions obtained from \autoref{eq:CbyC_z}. Points are the lattice data while the dashed lines are the theoretical expressions. The solid lines joining the points are only to guide the eye. The red horizontal line is the constant value corresponding to the asymptotic limit of $z\rightarrow\infty$. }
\label{fig:free_CbyC_z}
\end{figure}

According to \autoref{eq:gamma_sigma_free}, $\Gamma_\text{free}$ and $\Sigma_\text{free}$ approach quadratic and quartic polynomials respectively in the large-$\z$ limit. To understand how these limits are approached we look at $\Gamma_\text{free}/\z^2$ (\autoref{fig:free_CbyC_z} (left)) and $\Sigma_\text{free}/\z^4$ (\autoref{fig:free_CbyC_z} (right)). From \autoref{eq:gamma_sigma_free}, these are equal to
\begin{align} && 
\frac{\Gamma_\text{free}}{\z^2} = {- 4 + \frac{4}{\pi\z} -\frac{2}{\pi^2\z^2}}\,, &&
\frac{\Sigma_\text{free}}{\z^4} = {16 - \frac{32}{\pi\z} +\frac{16}{\pi^2\z^2}}\,. &&
\label{eq:CbyC_z}
\end{align}
The curves approach the value of the constant term, given in \autoref{eq:CbyC_z}, asymptotically as the contribution from other terms decreases at large $\z$. The asymptotic values are depicted by a horizontal red line in \autoref{fig:free_CbyC_z}. $\Gamma/\z^2$ approach the negative constant value of $-4$ from above while $\Sigma/\z^4$ approach a positive constant value of 16 from below. This asymptotic behavior is due to the different signs of the highest and second-highest coefficients. 

We also note that in both figures \autoref{fig:free_GS_isoscalar} and \autoref{fig:free_CbyC_z}, the $\z=N_\sigma/2N_\tau$ data point does not have exaggerated oscillations, unlike the nearby points. While the correlator and its derivatives deviate from the exact expression for this point in \autoref{fig:free_CbyC_isoscalar}(left), for the ratios $\Gamma$ and $\Sigma$, the last point seems to be unaffected by the boundary effect with the values matching the exact expression. We will see that the same features also appear in the finite-temperature data.

\section{Screening Correlators at Finite $\boldmath{T}$ and $\boldmath{\mS}$}
\label{sec:finite_T}
In this Section, we will present a new method of calculating the second-order correction to the $\mS=0$ pseudoscalar screening mass at finite temperature. Ignoring terms of $\O(e^{-4\pi\z})$ and higher in \autoref{eq:scr_corr_free}, we see that for $\mS=0$, the free theory correlator can be written as
\begin{align}
  &\frac{C(z,\mS=0)}{T^3} = Ae^{-Mz} \quad \text{where} \notag \\
  &A = \frac{3}{2\z}\left(1+\frac{1}{2\pi\z}\right) \quad \text{and} \quad M = 2 \pi T,
\end{align}
are the free theory screening amplitude and screening mass. For $\mS\ne0$, the free theory correlator can still be written as $Ae^{-Mz}$ provided we allow $A$ and $M$ to take complex values i.e.
\begin{align}
    &\frac{C_\text{free}(z,T,\mS)}{T^3} = \text{Re}\Big[A(\mS)e^{-zM(\mS)}\Big], \notag \\
    &\phantom{\frac{C_\text{free}(z,T,\mS)}{T^3}} = e^{-zM_R(\mS)}\Big[A_R(\mS)\cos(zM_I(\mS)) + A_I(\mS)\sin(zM_I(\mS))\Big],\notag \\
    &M(\mS) = 2\pi T + 2i\mS \equiv M_R(\mS) + iM_I(\mS), \notag \\
    &A(\mS) = \frac{3}{2zT}\left(1+\frac{1}{2\pi zT}\right)\left(1 - i\frac{\mS}{\pi T}\right) \equiv A_R(\mS) - iA_I(\mS).
\label{eq:finite_muS_correlator}
\end{align}

Similar to \autoref{eq:finite_muS_correlator}, we postulate that the finite-temperature screening correlator can be written with complex screening mass and screening amplitude as
\begin{equation} 
    \frac{C(z,T,\mS)}{T^3} = e^{-zM_R(\mS)}
    \Big[A_R(\mS)\cos\left(zM_I(\mS)\right) + A_I(\mS)\sin\left(zM_I(\mS)\right)\Big].
\label{eq:scr_corr_finiteT}
\end{equation}
$M_R$ and $A_R$ are even functions of $\mS$, while $M_I$ and $A_I$ are odd functions of $\mS$. This can be seen by looking at the hermitian conjugate of $G$ (\autoref{eq:meson_operator}) and $\gamma_5-$hermiticity of $\Delta$ (\autoref{eq:determinant})
\begin{eqnarray}
    G(x,\mS)^*=G(x,-\mS)&,\,\,\,\,\,\,&\Delta(T,\mS)^*=\Delta(T,-\mS).
\end{eqnarray}
Taking hermitian conjugate of \autoref{eq:pion_corr} and using the reality of the screening correlator \autoref{eq:scr_corr_finiteT}, we see that the screening correlator must be an even function of $\mS$
\begin{eqnarray}
    C(x,T,\mS)=C(x,T,-\mS)
\end{eqnarray}
This requires that the odd (even) derivatives of $M_R$ and $A_R$ (of $M_I$ and $A_I$) vanish at $\mS=0$. By successively differentiating \autoref{eq:scr_corr_finiteT} w.r.t. $\hm$, we obtain (primes denote differentiation w.r.t. $\hm$ at $\hm=0$):
\begin{align}
\Gamma(z) &= \frac{A_R''}{A_R} 
+ z\left[2\frac{A_I'}{A_R}M_I' - M_R''\right]
- z^2\left(M_I'\right)^2, \notag \\
 &\equiv \alpha_2\z^2+\alpha_1\z+\alpha_0, \label{eq:Gamma}\\
\Sigma(z) &= \frac{A_R''''}{A_R} + z\left[4\frac{A_I'}{A_R}M_I'''+4\frac{A_I'''}{A_R}M_I'-M_R''''-6M_R'' \frac{A_R''}{A_R}\right] \notag\\
&+z^2\left[3M_R''^2-12\frac{A_I'}{A_R}M_I'M_R''-4M_I'M_I'''-6 M_I'^2\frac{A_R''}{A_R}\right]
\notag\\
&+z^3\left[6M_R''M_I'^2-4\frac{A_I'}{A_R}M_I'^3\right]+z^4\left(M_I'\right)^4,
\notag \\
&\equiv\beta_4\z^4+\beta_3\z^3+\beta_2\z^2+\beta_1\z+\beta_0.  \label{eq:Sigma}
\end{align}

\autoref{eq:Gamma} and \autoref{eq:Sigma} are then quadratic and quartic polynomials in $\z$ for $\Gamma(z)$ and $\Sigma(z)$ respectively, just as for the free theory (\autoref{eq:gamma_sigma_free}). The lowest order corrections $M_I'$ and $M_R''$ to the screening mass can be obtained from the coefficients of these polynomials as
\begin{align} && && &&
\hat{M}_I'=\left(-\alpha_2\right)^{1/2}=\beta_4^{1/4} && \text{and} &&
\hat{M}_R''=\frac{1}{4}\left(2\alpha_1-\frac{\beta_3}{\alpha_2}\right). && && &&
\label{eq:mr2}
\end{align}
where $\hat{M}=M/T$. Substituting the free theory values for these coefficients from \autoref{eq:gamma_sigma_free} in the above equations, we obtain the correct values $\hat{M}_I'\text{(free theory)}=2$ and $\hat{M}_R''\text{ (free theory)}=0$.

In deriving \autoref{eq:Gamma} and \autoref{eq:Sigma}, we only considered contributions from a single state. Previously, we had derived equations similar to the above equations, but additionally including the contributions from the first excited state as well~\cite{Thakkar:2022frk}. The excited states contribute only at shorter distances $\z \lesssim 2.25$. Considering the contribution of the excited states was necessary due to the large number of fit parameters, which necessitated going to smaller $\z$. By contrast, in this paper, we present a new method that allows us to fit the data at larger $\z$ by reducing the number of fit parameters and thus decreasing the uncertainty in the fit parameters. Hence in this paper, we will work only with \autoref{eq:Gamma} and \autoref{eq:Sigma} and not consider the contributions coming from the excited states.

\subsection{Finite Temperature Analysis}
\par Our finite temperature analysis was carried out keeping the temporal extent on the lattice fixed at $N_\tau=8$. The analysis was done for two temperatures viz. $T=2.24$ GeV and $T=2.90$ GeV, and for two volumes $N_\sigma=32$ and $N_\sigma=64$. The number of configurations analyzed and quark masses for each $\beta$ are listed in \autoref{tab:conf_list}. The temperature was determined from the relation $T=1/\Nt a(\beta)$, where $a(\beta)$ is the lattice spacing at a given gauge coupling $\beta$, using the updated parametrization for the function $af_\text{K}(\beta)$ given in Ref.~\cite{Bazavov:2019www}. The strange quark mass $m_s$ was set to its physical value using the parametrization provided in Ref.~\cite{Bazavov:2014pvz}. The light quark mass $m_\ell$ was set to $m_s/20$, corresponding to a nearly physical pion mass of 160 MeV at $T=0$.

The gauge configurations were generated using the (2+1)-flavor HISQ/tree action~\cite{Follana:2006rc,MILC:2010pul,Bazavov:2011nk}. The Bielefeld RHMC GPU code was used to generate the configurations~\cite{Bollweg:2021cvl}. The configurations were generated using the leapfrog evolution with molecular dynamics step size 0.2 and trajectory length of 5 steps keeping the acceptance rate between 65\% and 80\%, with every $10^\text{th}$ configuration being saved.

On each saved configuration, we calculated the derivatives of the correlator, which we call correlator-like operators, as well as the derivatives of the fermion determinant, which we call the trace-like operators in \autoref{app:isoscalar_chempot}. The former were calculated by using 8 point sources per configuration, which we placed at $n_i=0$ or $N_\sigma/2$ for $i\in\{x,y,z\}$ keeping $n_t=0$. The latter were estimated stochastically using 1000 Gaussian noise vectors per configuration.

\begin{table}[!h]
	\centering
	\begin{tabular}{ |c|c|c|c|c| } 
		\hline
		$\beta$ & $T$ [GeV]& $N_\sigma$ & $m_s$ & configurations \\ 
		\hline
		9.670 & 2.90 & 32 & 0.002798& 12700\\ 
		     &      & 64 & 0.002798& \phantom{1}6000\\ \hline
		9.360 & 2.24 & 64 & 0.003691& \phantom{1}6000\\
		\hline
	\end{tabular}
	\caption{The list of configurations used for the finite temperature. All the configurations used here have $N_\tau=8$.}
	\label{tab:conf_list}
\end{table}

\begin{figure}[!t]
	 \centering
\includegraphics[width=.48\linewidth]{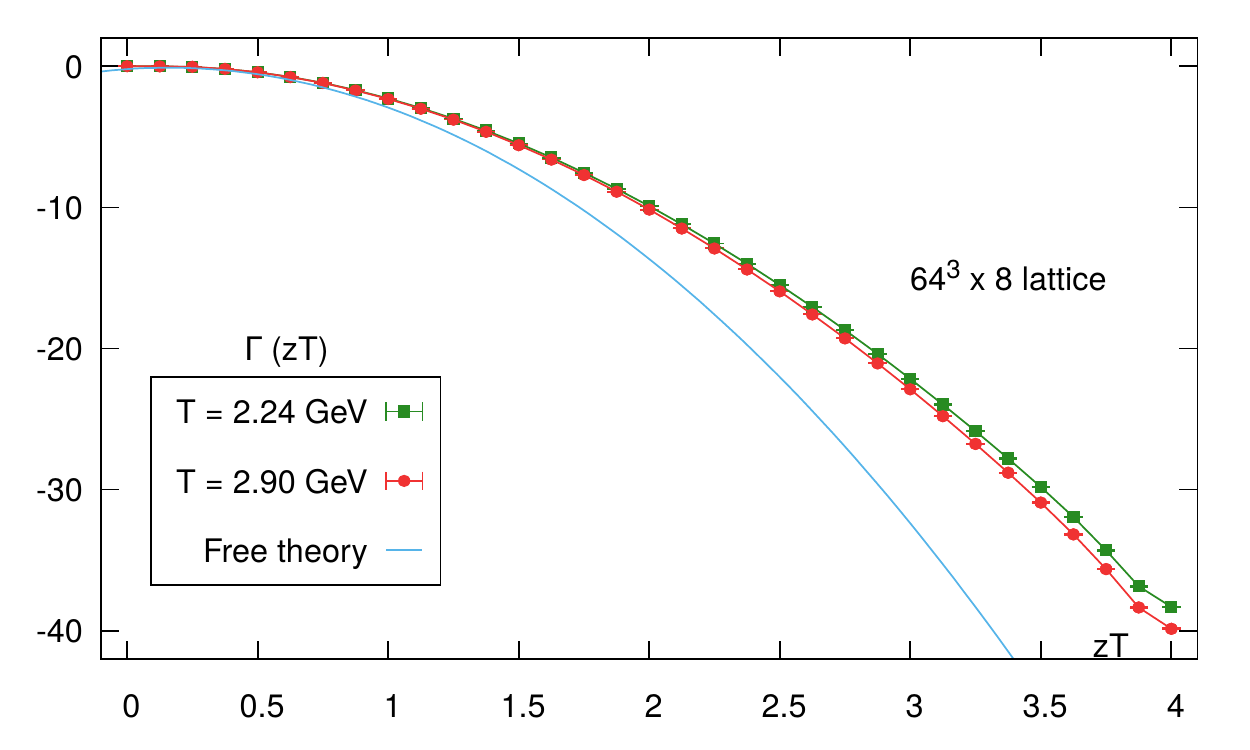} 
\includegraphics[width=.48\linewidth]{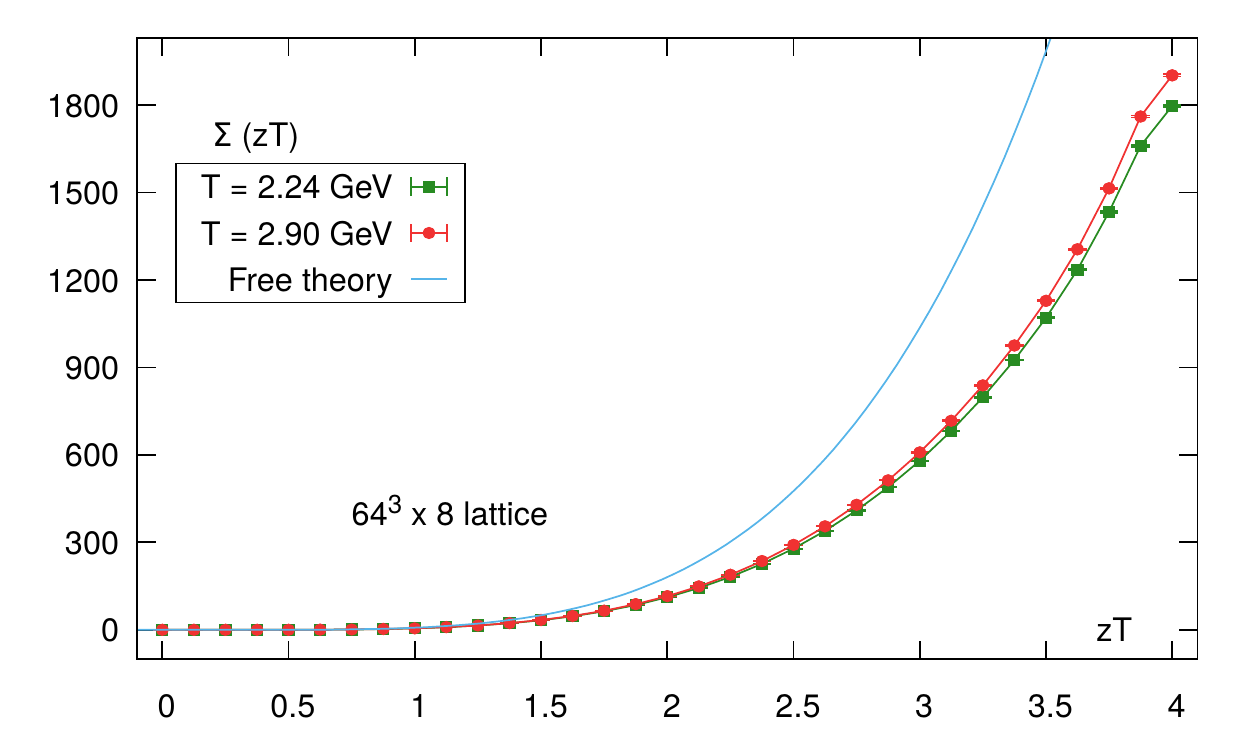} \\
    \includegraphics[width=.48\linewidth]{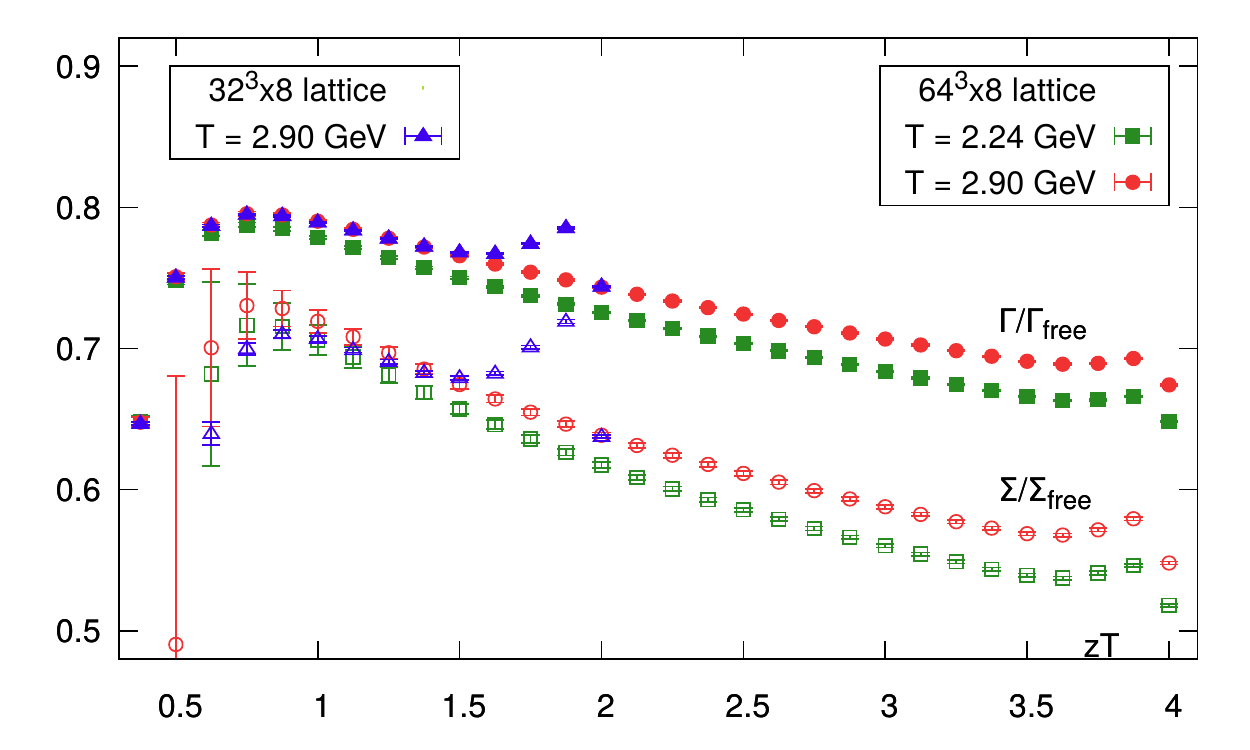}
	\caption{(Top left) $\Gamma$ and (top right) $\Sigma$ for T$=2.24$ GeV and T$=2.90$ GeV along with the free theory expression \autoref{eq:gamma_sigma_free} plotted against $n_z$ measured on $64^3\times8$ lattice. (Bottom) $\Gamma/\Gamma_\text{free}$ and $\Sigma/\Sigma_\text{free}$ plotted against $\z$ for the two temperatures as well as the two volumes given in \autoref{tab:conf_list}.}
	\label{fig:GS_isoscalar}
\end{figure}

We compare the finite temperature results for $\Gamma(\z)$ and $\Sigma(\z)$ to the corresponding free theory expressions in \autoref{fig:GS_isoscalar}~(top left) and \autoref{fig:GS_isoscalar}~(top right) respectively. In each plot, we plot the results for both $T=2.24$ GeV and $T=2.90$ GeV. Such high temperatures were considered as our ansatz was obtained by comparison with free theory expression and thus is more applicable at higher temperatures. Although both $\Gamma(\z)$ and $\Sigma(\z)$ at finite temperatures show a polynomial-like behavior similar to the corresponding free theory expressions, they differ by as much as 30-40\% from the corresponding free theory correlators, with the fourth derivative $\Sigma$ being comparatively farther from the free theory limit than the second derivative $\Gamma$. Furthermore, both $\Gamma$ and $\Sigma$ seem to approach the corresponding free theory results as the temperature is increased, although the approach to the free theory is very slow. We note that a similar slow approach to the free theory limit has also recently been observed in the case of the zero chemical potential screening masses~\cite{DallaBrida:2021ddx}. Thus we expect the polynomial coefficients in \autoref{eq:Gamma} and \autoref{eq:Sigma} to differ significantly from their free theory values (\autoref{tab:fit_free}). This is seen more clearly in \autoref{fig:GS_isoscalar} (bottom), in which we plot $\Gamma/\Gamma_\text{free}$ and $\Sigma/\Sigma_\text{free}$ given by
\begin{align} && 
\frac{\Gamma}{\Gamma_\text{free}} = \frac{\alpha_2\z^2+\alpha_1\z+\alpha_0}{- 4\z^2 + 4\z/\pi -2/\pi^2}\,, &&
\frac{\Sigma}{\Sigma_\text{free}} = \frac{\beta_4\z^4+\beta_3\z^3+\beta_2\z^2+\beta_1\z+\beta_0}{16\z^4 - 32\z^3/\pi + 16\z^2/\pi^2}\,, &&
\label{eq:CbyCfree}
\end{align}
as functions of $\z$ for all the temperatures and volumes given in \autoref{tab:conf_list}. For large $\z$ however, both ${\Gamma}/{\Gamma_\text{free}}$ and ${\Sigma}/{\Sigma_\text{free}}$ curves are seen to slowly decrease to approach the corresponding asymptotic values of $-\alpha_2/4$ and $\beta_4/16$ according to \autoref{eq:CbyCfree}. The deviation from the free theory value of unity indicates that the finite temperature polynomial coefficients differ significantly from the free theory ones. 

Based on our results for the two volumes considered for $T=2.90$ GeV, we conclude that there are no significant finite-volume effects for these temperatures. However, both $\Gamma/\Gamma_\text{free}$ and $\Sigma/\Sigma_\text{free}$ curve upwards for all the data sets as $\z\rightarrow\Ns/(2\Nt)$, indicating the presence of boundary effects. These boundary effects do not seem to affect the point at $z=\Ns/2$. The upward deviation starts at around $\z=1.5$ for $\Ns=32$ and around $\z_{\text{max}}=3.5$ for $\Ns=64$. Due to this, we set an upper limit of $\z=3.25$ for our fits on the $\Ns=64$ lattices while also however including the $\z=\Ns/(2\Nt)$ point.

\begin{figure}[!t]
	\centering
\includegraphics[width=.48\linewidth]{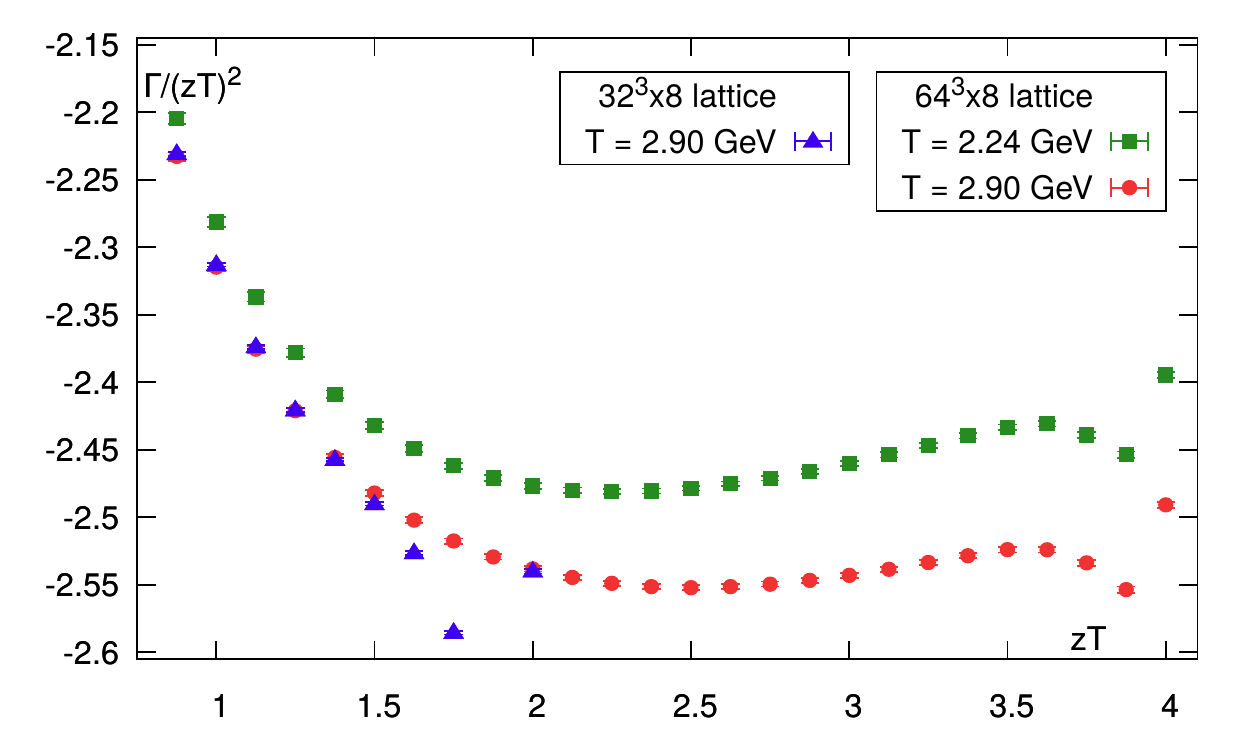}
\includegraphics[width=.48\linewidth]{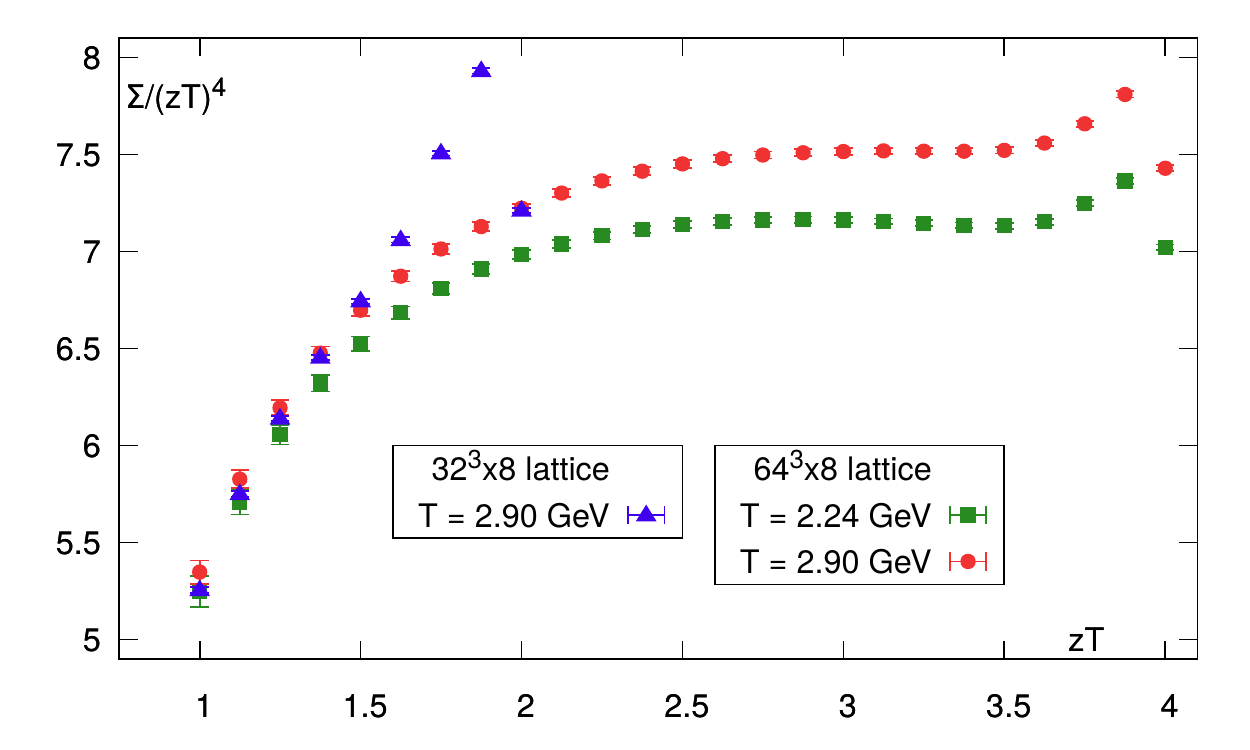}
	\caption{(left) $\Gamma/\z^2$ and (right) $\Sigma/\z^4$ data plotted against $n_z$ for the two temperatures as well as the two volumes given in \autoref{tab:conf_list}.}
	\label{fig:GS_byz}
\end{figure}

Similar to the free theory, we plot $\Gamma/\z^2$ and $\Sigma/\z^4$ in \autoref{fig:GS_byz} (left) and \autoref{fig:GS_byz} (right) respectively to understand the approach of $\Gamma$ and $\Sigma$ to their respective asymptotic limits. We rewrite \autoref{eq:Gamma} and \autoref{eq:Sigma} as
\begin{align} && 
\frac{\Gamma}{\z^2} = -{|\alpha_2| - \frac{|\alpha_1|}{\z} + \frac{\alpha_0}{\z^2}} && \text {and} &&
\frac{\Sigma}{\z^4} = {\beta_4 + \frac{\beta_3}{\z} -\frac{|\beta_2|}{\z^2}} + \O\left(\frac{1}{\z^3}\right)\,, && 
\label{eq:GS_z}
\end{align}
The boundary effects mentioned earlier are observable near $\z=N_\sigma/2N_\tau$ as the data points curve downwards for both temperatures although as noticed earlier, the $\z=N_\sigma/2N_\tau$ point seems unaffected by the boundary effects. While the data do seem to approach a constant value asymptotically, however unlike the free theory, we observe that the finite temperature curves attain a minimum for $\Gamma/\z^2$ and a maximum for $\Sigma/\z^4$. In anticipation of this behavior, we have made $\alpha_2$, $\alpha_1$ and $\beta_2$ in \autoref{eq:GS_z} negative. By comparing the results for the two temperatures, we see that these minima and maxima shift to larger $\z$ as the temperature is increased, with $\z\to\infty$, i.e. vanishing, as $T\to\infty$. 

\begin{figure}[!t]
	\centering
\includegraphics[width=.48\linewidth]{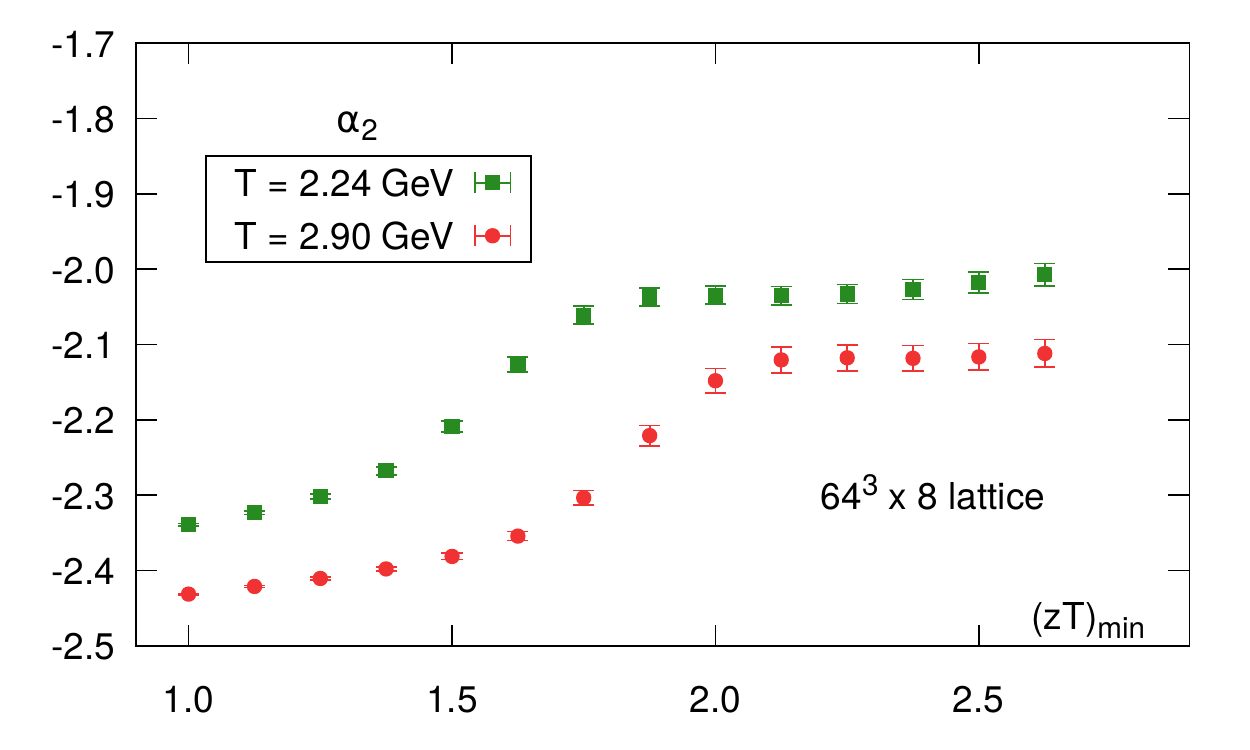}
\includegraphics[width=.48\linewidth]{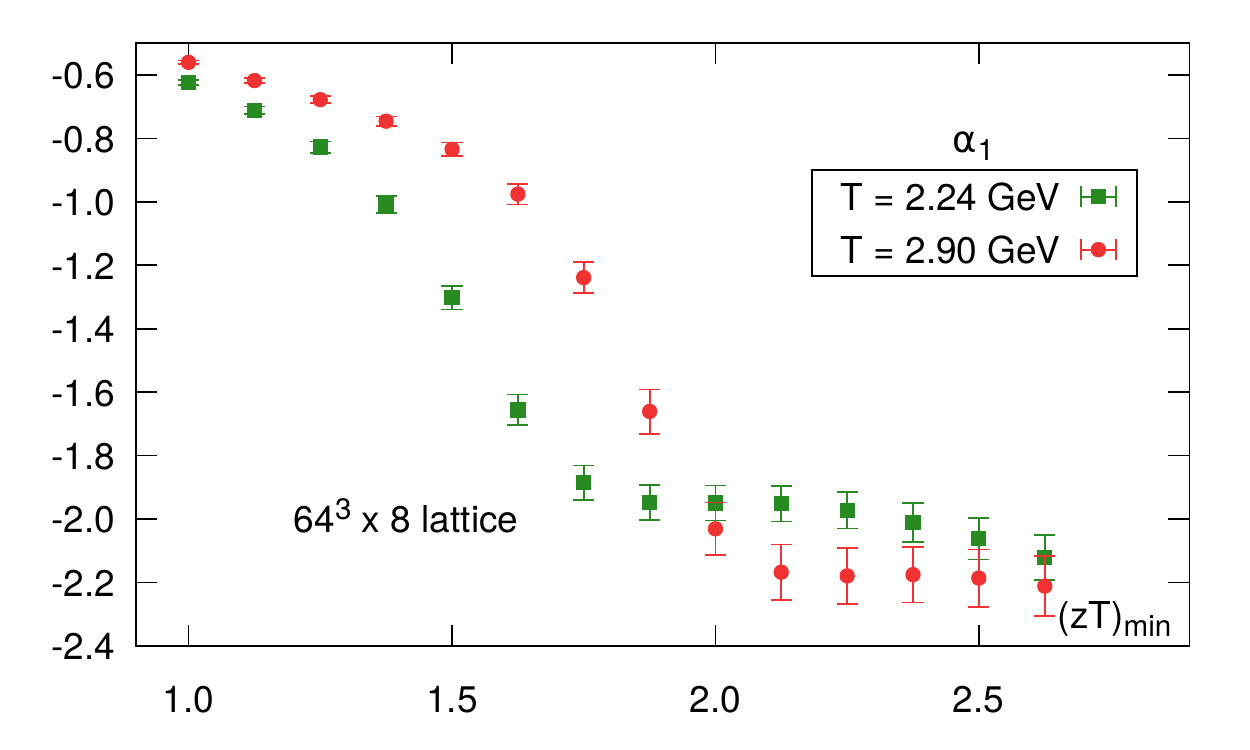}\\
\includegraphics[width=.48\linewidth]{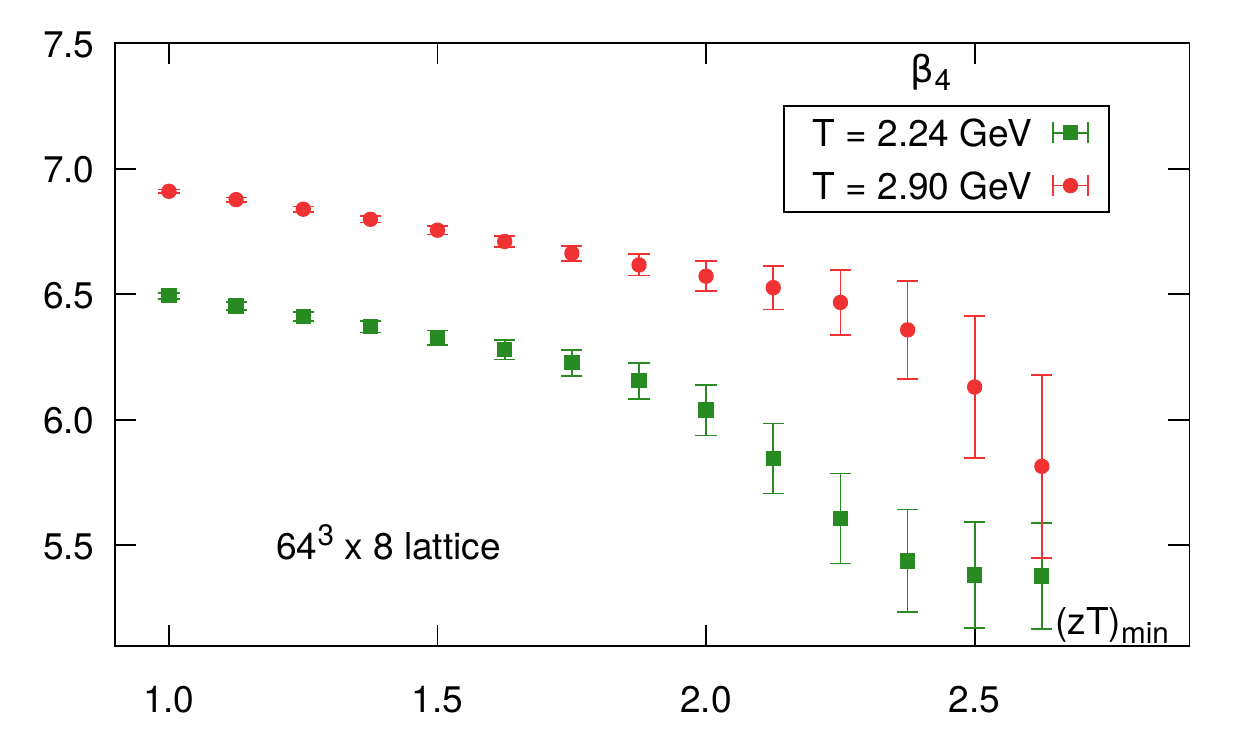}
\includegraphics[width=.48\linewidth]{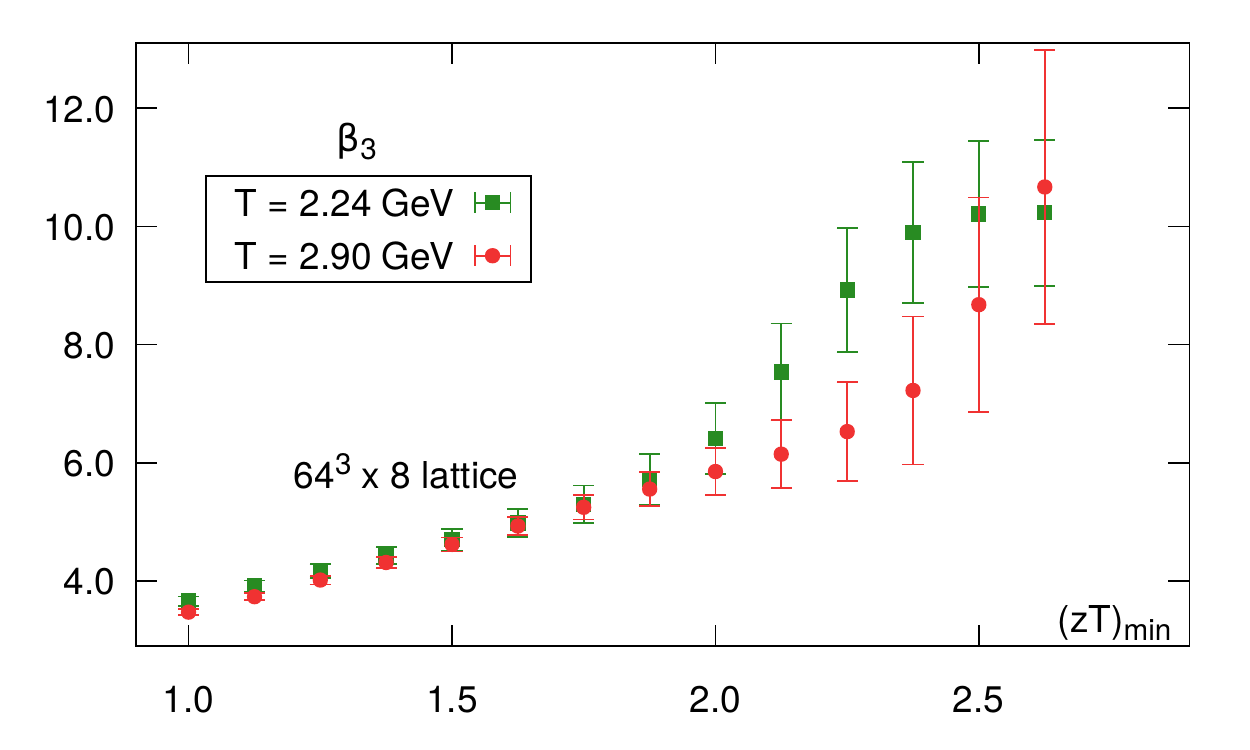}
	\caption{(top left) $\alpha_2$, (top right) $\alpha_1$,(bottom left) $\beta_4$, and (bottom right) $\beta_3$ coefficients obtained using fitting the $\Gamma/\z^2$ and $\Sigma/\z^4$ respectively for T=2.24 GeV and T=2.90 GeV on $64^3\times8$ lattice. The upper window is fixed to $\z_\text{max}=3.25$ while the lower window $\z_\text{min}$ is varied.}
	\label{fig:ab}
\end{figure}

From \autoref{eq:GS_z}, we can identify the location of the extrema of $\Gamma/\z^2$ and $\Sigma/\z^4$ as
\begin{align} && 
\z_\Gamma = -2 \,\frac{\alpha_0}{\alpha_1}\,, &&
\z_\Sigma = -2 \,\frac{\beta_2}{\beta_3}\,. &&
\label{eq:zGamma_zSigma}
\end{align}
Instead of keeping all three coefficients as fit parameters, the extremum points $\z_\Gamma$ and $\z_\Sigma$ were located for each jackknife sample using spline fittings, and their values were used to re-express the lowest order coefficients $\alpha_0$ and $\beta_2$ in terms of $\alpha_1$ and $\beta_3$. Reducing the number of fit parameters from three to two resulted in better fits to the data. The fits were carried out for various fit windows $[\z_\text{min},\z_\text{max}]$. The fit results showed very little variation for changing $\z_\text{max}$ and thus the upper fit window was fixed to $\z_\text{max}=3.25$. Subsequently, we sought to obtain stable results for the fit coefficients by varying $\z_\text{min}$. Our results for $\alpha_1$, $\alpha_2$, $\beta_4$, and $\beta_3$ are plotted in \autoref{fig:ab}. We note that our procedure yields stable values for $\alpha_1$ and $\alpha_2$, whereas the plateaus for $\beta_4$ and $\beta_3$ are not reached for $T = 2.90$ GeV. We need to consider larger lattices in order to get more reliable values of these coefficients, especially at higher temperatures. For our estimates of $\beta_4$ and $\beta_3$ at $T = 2.90$ GeV, we approximated their values from fits with windows keeping $\z_\text{min}=2.625$.

\begin{table}[!b]
{\centering
\begin{tabular}{|c|c|c|c|c|c|c|} \hline
 Temp (T)&  $\z_\Gamma$ & $\alpha_2$ & $\alpha_1$  & $\z_\Sigma$ &  $\beta_4$ & $\beta_3$ \\ \hline
2.24 GeV
&  2.269(23) & -2.034(13) & -1.955(57) & 2.860(50) & 5.383(218) & 10.091(1255)  \\ \hline
2.90 GeV
& 2.500(16) & -2.117(18) & -2.175(87)   & 3.125(25) & 5.815(365) & 10.667(2321)  \\\hline
Free th. & & $-4$ & $4/\pi\approx1.273$ &  & 16 & $-32/\pi\approx-10.186$  \\ \hline 
\end{tabular}
 \begin{center}
\begin{tabular}{|c|c|c|c|} \hline
 Temperature & $\hat{M}_R  (\hm=0)$ &$\hat{M}_R''$ & $\hat{M}_I'$\\ \hline
2.24 GeV&
6.337(1)& 0.263(169) & 1.426(5)\\ \hline
2.90 GeV
& 6.352(1) & 0.172(328) & 1.455(6) \\\hline
Free theory& $2\pi\approx6.283$ & 0 & 2 \\ \hline  
\end{tabular}
\end{center}
\caption{Values listed for extremum point $\z_\Gamma (\z_\Sigma)$ for the function $\Gamma/\z^2 (\Sigma/\z^4)$ along with the obtained values for fit parameters $\alpha_2$ and $\alpha_1$ ($\beta_4$ and $\beta_3$) for two temperatures on lattices with volume $64^3\times8$. Two-state fit screening mass $\hat{M}_R (\hm=0)$, along with $\hat{M}_R''$ and $\hat{M}_I'$ obtained using \autoref{eq:mr2} are listed for the two temperatures obtained on lattices with volume $64^3\times8$.}
\label{tab:data_final}
}
\end{table}

By fitting a constant value to the plateaus, we obtain the best-fit values for each coefficient, which we present in \autoref{tab:data_final}. In quoting the errors on these fits, we note that there are two separate sources of error. In the first place, the fitter itself returns an error on each fit parameter. Second, there is also the variation of the fit parameter itself from one jackknife sample to the next. Since we found that the former was always much smaller than the latter, we have only quoted the jackknife errors in \autoref{tab:data_final}. 

 We tabulate our results for all the fit parameters, namely $\z_\Gamma$, $\z_\Sigma$, $\alpha_1$, $\alpha_2$, $\beta_3$ and $\beta_4$, in \autoref{tab:data_final}. From the Table, we see that the highest coefficients $\alpha_2$ and $\beta_4$ have the same sign for both the free theory and finite temperature. The second highest coefficients $\alpha_1$ and $\beta_3$ have different signs for the free theory and finite temperature, as we have already seen earlier. We also note that all the coefficients are far from the free theory values, although they seem to slowly approach the free theory as the temperature is increased.

 By substituting the results for the fit parameters into \autoref{eq:mr2}, we obtain the lowest order corrections to the screening mass, namely $M_I'$ and $M_R''$. We have also listed our results for both these quantities in \autoref{tab:data_final}. In the same Table, we have also listed the values of the screening masses $M_R(0)$ obtained for each temperature. These were calculated by fitting the pseudoscalar correlator to one-state and two-state fits and using the Akaike Information Criterion (corrected) (AICc) to obtain the mass plateau from the two-state fit. The procedure is identical to that used in Ref.~\cite{Bazavov:2019www} for the calculation of the $\mu=0$ screening masses, and we refer the reader to that paper and the references therein for further details.
 
We find that the screening mass $M_R(0)$ values are larger than the free theory value of $2\pi$ at the temperatures of our analysis. We also see that $M_R''(0)$ is around 4\% of $M_R(0)$ for both temperatures. Assuming higher-order corrections to be negligible, this suggests that $M_R(\mS)$ differs from $M_R(0)$ by only about 2\% for $\hm=1$ near $T\sim2.5$~GeV. Note however that our results for $M_I'(0)$ differ by as much as 25-30\% from the free theory value of 2. Both $M_R''(0)$ and $M_I'(0)$ seem to approach the respective free theory values with increasing temperature, although the rate of approach is very slow. While this is surprising, we note that a recent determination of the pseudoscalar screening mass at $\mu=0$ too found a similar exponentially slow approach to the free theory value~\cite{DallaBrida:2021ddx}.

\section{Conclusions}
\label{sec:conclusions}
In this paper, we introduced a new way to calculate the finite-$\mS$ corrections to the pseudoscalar screening mass. Our approach is based on the method of Taylor expansions, in which both the screening correlator and the screening mass are expanded in a Taylor series in $\hm\equiv\mS/T$.  In the free theory, the finite density screening correlator manifests oscillatory behavior in addition to decaying exponentially with increasing separation~\cite{vepsalainen2007mesonic}. In this work, we showed that these oscillations can be taken into account by making the screening mass complex viz. $M=M_R+iM_I$. The real part $M_R(\mS)$ is the familiar screening mass in the $\mS\to0$ limit while the imaginary part $M_I$ vanishes in the same limit. However, the imaginary part is responsible for the oscillations at $\mS\neq0$ and hence it is necessary to incorporate it into our formalism in order to obtain a reliable estimate of $M_R''(0)$ at finite temperature.

We expanded the free correlator in a Taylor series in $\hm$ and calculated the first four Taylor coefficients on an $80^3\times8$ lattice using HISQ fermions. Our results showed very good agreement with the theoretical expressions \autoref{eq:taycoeff_free}, down to $\z\sim0.3$. By combining the results of different orders, we were also able to show that the correlator displays the expected oscillations as a function of $\hm$ (\autoref{fig:free_CbyC_isoscalar}). We also showed that the ratios of Taylor coefficients $\Gamma(\z)$ and $\Sigma(\z)$ should behave as quadratic and quartic polynomials respectively in the large-$\z$ limit (\autoref{eq:gamma_sigma_free}). Our fits to the free theory results confirmed the expected behavior, while also indicating that one had to go to $\z\gtrsim2$-3 to observe this asymptotic behavior. We also showed how the screening mass corrections $M_R''(0)$ and $M_I'(0)$ could be extracted from the coefficients of these polynomials (\autoref{eq:mr2}). Finally, we applied our formalism to screening correlators at two temperatures $T=2.24$~GeV and $T=2.90$~GeV calculated on $64^3\times8$ lattices with the HISQ/tree action. We extracted results for $M_R''(0)$ and $M_I'(0)$ for both these temperatures. In both cases, the screening mass correction $M_R''(0)$ was positive and around 4\% of $M_R(0)$. We also found significant differences from the free theory values, for all quantities but especially $M_I'(0)$, and a very slow approach to the free theory as the temperature was increased.

\acknowledgments
We thank Frithjof Karsch, Anirban Lahiri, Sourendu Gupta, Saumen Datta, and Rishi Sharma for helpful discussions and suggestions. The data for this project were generated at the Centre for High Energy Physics of the Indian Institute of Science, Bengaluru. The Bielefeld RHMC GPU code was used to generate the gauge configurations, and a modification of the same code was used in calculating the Taylor coefficients.

\bibliographystyle{apsrev4-1.bst}
%

\appendix

\section{Third and Fourth Derivatives of $\langle\langle\mathrm{tr}\big[P(\x,0,\mS) P(0,\x,\mS)\big]\rangle\rangle$}
\label{app:corr_derivatives}
The first two derivatives have already been given in \autoref{eq:Ck_zero_to_two}. The third derivative is equal to
\begin{align}
\G^{(3)}(\x,T) &=
\left\langle G''' \right\rangle + 3 \left\langle G'' \frac{\Delta'}{\Delta}\right\rangle
+ 3 \left\langle G' \frac{\Delta''}{\Delta}\right\rangle+  \left\langle G \frac{\Delta'''}{\Delta}\right\rangle \notag \\
&-3\left\langle \frac{\Delta''}{\Delta}\right\rangle \left(  \left\langle G' \right\rangle+  \left\langle G \frac{\Delta'}{\Delta}\right\rangle \right),
\end{align}
while the fourth derivative is given by
\begin{align}
\G^{(4)}(\x,T) &=
\left\langle G'''' \right\rangle + 4 \left\langle G''' \frac{\Delta'}{\Delta}\right\rangle
+ 6 \left\langle G'' \frac{\Delta''}{\Delta}\right\rangle\notag\\&+ 4 \left\langle G' \frac{\Delta'''}{\Delta}\right\rangle+ \left\langle G \frac{\Delta''''}{\Delta}\right\rangle +6\left\langle G \right\rangle \left\langle \frac{\Delta''}{\Delta}\right\rangle^2 \notag \\
 &- \left( 6 \left\langle G'' \right\rangle+ 12 \left\langle G' \frac{\Delta'}{\Delta}\right\rangle+ 6 \left\langle G \frac{\Delta''}{\Delta}\right\rangle \right)\left\langle \frac{\Delta''}{\Delta}\right\rangle \notag\\&-  \left\langle G \right\rangle \left\langle \frac{\Delta''''}{\Delta}\right\rangle.
\end{align}
Just as for \autoref{eq:Ck_zero_to_two}, we have set terms containing $\langle \Delta' \rangle$, $\langle \Delta''' \rangle$, etc. to zero since these vanish by the $CP$ symmetry of the QCD action.

\section{Derivatives of $G$ and $\Delta$}
\label{app:isoscalar_chempot}
In this Appendix, we present the formulas for the first four $\mS$ derivatives of $G(\x,T,\mS)$ (\autoref{eq:meson_operator}) and $\Delta(T,\mS)$ (\autoref{eq:determinant}) in terms of various operators. We will refer to the operators appearing in the derivatives of $G(\x,T,\mS)$ as correlator-like operators and $\Delta(T,\mS)$ as trace-like operators.

The calculation of the derivatives proceeds from the following starting point (all arguments are now made implicit):
\begin{equation}
    P'=-PM'P.
\end{equation}
The formulas presented here are specific to the staggered case with degenerate $u$ and $d$ quarks. The corresponding expressions for non-staggered fermions can be looked up in Ref.~\cite{QCD-TARO:2001lhr}.

\subsection{Correlator-like operators}
\begin{subequations}
\begin{flalign}
G'= &\,\, -2\,i\,\Im\, \Tr \left[ (PM'P)
P^{\dagger}\right]&
\end{flalign}
\begin{flalign}
G'' =&\,\, 4\,\Re\,\Tr\, \left[ (PM'PM'P)
P^{\dagger} \right] - 2\,\Re\,\Tr\, \left[ (PM''P)
P^{\dagger}  \right]\nonumber &\\&
 -2\,\Tr \left[ (PM'P)
(PM'P)^{\dagger}  \right]& 
\end{flalign}
\begin{flalign}
G'''=&\,\, -12\,i\,\Im\,\Tr\, \left[ (PM'PM'PM'P)
P^{\dagger}  \right] + 6 \,i\,\Im\,\Tr\, \left[ (PM'PM''P)
P^{\dagger}  \right]\nonumber & \\& + 6 \,i\,\Im\,\Tr\, \left[ (PM''PM'P)
P^{\dagger}  \right] - 6 \,i\,\Im\,\Tr\, \left[ (PM''P)
(PM'P)^{\dagger}  \right]\nonumber & \\& -2 \,i\,\Im\,\Tr\, \left[ (PM'''P)
P^{\dagger}  \right]+12\,i\,\Im\Tr \left[ (PM'PM'P)
(PM'P)^{\dagger}  \right]&
\end{flalign}
\begin{flalign}
G''''=& \,\,48\,\Re\,\Tr\, \left[ (PM'PM'PM'PM'P)
P^{\dagger}  \right] -48 \,\Re\,\Tr\, \left[ (PM'PM'PM'P)
(PM'P)^{\dagger}  \right]\nonumber & \\ & + 24 \,\,\Tr\, \left[ (PM'PM'P)
(PM'PM'P)^{\dagger}  \right]  -24 \,\Re\,\Tr\, \left[ (PM''PM'PM'P)
P^{\dagger}  \right]\nonumber   & \\ & -24 \,\Re\,\Tr\, \left[ (PM'PM''PM'P)
P^{\dagger}  \right] -24 \,\Re\,\Tr\, \left[ (PM'PM'PM''P)
P^{\dagger}  \right]\nonumber  & \\
& +24 \,\Re\,\Tr\, \left[ (PM''PM'P)
(PM'P)^{\dagger}  \right]+24 \,\Re\,\Tr\, \left[ (PM'PM''P)
(PM'P)^{\dagger}  \right]\nonumber  &\\& -24 \,\Re\,\Tr\, \left[ (PM''P)
(PM'PM'P)^{\dagger}  \right] + 8 \,\Re\,\Tr\, \left[ (PM'''PM'P)
P^{\dagger}  \right]\nonumber   &\\ & +8\,\Re\,\Tr\, \left[ (PM'PM'''P)
P^{\dagger}  \right] -8\,\Re\,\Tr \left[ (PM'''P)
(PM'P)^{\dagger}  \right]\nonumber & \\& +12\,\Re\,\Tr\, \left[ (PM''PM''P)
P^{\dagger}  \right]+6\,\Tr \left[ (PM''P)
(PM''P)^{\dagger}  \right]\nonumber & \\& -2\,\Re\,\Tr\, \left[ (PM''''P)
P^{\dagger}  \right]
\end{flalign}
\end{subequations}
\subsection{Trace-like operators}
\label{ssec:trace}
\begin{subequations} 
\begin{flalign}
\frac{\Delta'}{\Delta}&=\frac{1}{2} \Tr [M'P]&
\label{eq:sample_tr}
\end{flalign}
\begin{flalign}
\frac{\Delta''}{\Delta}
=&\,\,\frac{1}{2} \Tr [M''P]-\frac{1}{2} \Tr [M'PM'P]+\frac{1}{4} \left(\Tr [M'P]\right)^2&
\end{flalign}
\begin{flalign}
\frac{\Delta'''}{\Delta}
=&\,\,\Tr [M'PM'PM'P]-\frac{3}{2} \Tr [M''PM'P]+\frac{1}{2} \Tr [M'''P]\nonumber&\\
& +\frac{3}{4}\Tr [M'P]\Big\{\Tr [M''P]- \Tr [M'PM'P]+\frac{1}{6} \left(\Tr [M'P]\right)^2\Big\}&
\end{flalign}
\begin{flalign}
\frac{\Delta''''}{\Delta}
=&\,\,-3\Tr [M'PM'PM'PM'P]+6 \Tr [M''PM'PM'P]\nonumber&\\
&-2 \Tr [M'''PM'P]-\frac{3}{2} \Tr [M''PM''P]+\frac{1}{2} \Tr [M''''P] \nonumber&\\
&+ \Tr [M'P]\Big\{2\Tr [M'PM'PM'P]-3 \Tr [M''PM'P]+  \Tr [M'''P]+\frac{1}{16}\left(\Tr [M'P]\right)^3\Big\}\nonumber&\\
&  +\frac{3}{4}\Big\{ \Tr [M''P]- \Tr [M'PM'P]\Big\}\Big\{ \Tr [M''P]- \Tr [M'PM'P]+\left(\Tr [M'P]\right)^2\Big\}&
\end{flalign}
\end{subequations}
\end{document}